# Why cells grow and divide? General growth mechanism and how it defines cells' growth, reproduction and metabolic properties

YURI K. SHESTOPALOFF

We consider a general growth mechanism, which acts at cellular level and above (organs, systems and whole organisms). Using its mathematical representation, the growth equation, we study the growth and division mechanisms of *amoeba* and fission yeast *Schizosaccharomyces pombe*. We show how this mechanism, together with biomolecular machinery, governs growth and reproduction of cells, and these organisms in particular. This mechanism provides revealing answers to fundamental questions of biology, like why cells grow and divide, why and when cells' growth stops. It also sheds light on questions like why and how life originated and developed. Solving the growth equation, we obtain analytical expression for the growth curve of fission yeast as a function of geometrical characteristics and nutrient influxes for RNA and protein synthesis, and compare the computed growth curves with 85 experiments. Statistical evaluation shows that these growth curves correspond to experimental data significantly better than all previous approximations. Also, using the general growth mechanism, we show how metabolic characteristics of cells, their size and evolutionary traits relate, considering fission yeast. In particular, we found that fission yeast *S. pombe* consumes about 16-18 times more nutrients for maintenance needs than for biomass synthesis.

*Keywords: General growth law; amoeba; S. pombe; growth curve; biomass synthesis; maintenance costs; nutrients; metabolism; fundamental growth mechanisms*

## 1. Introduction

The subject of what kind of mechanisms regulate growth and division of cells and unicellular organisms has been intensively researched from different perspectives for several decades. Historically, it started from study of cell populations and single cells. A comprehensive review of both experimental and conceptual aspects of these developments can be found in Ref. 1.

Studies related to growth and division mechanisms can be roughly divided into the following broad categories. First explores biochemical mechanisms that support growth and reproduction. Works done in P. Nurse's Lab, by S. Martin, J. Tyson contributed much to understanding of particular biochemical mechanisms.[2-5] These studies concentrate on biochemical reactions which support growth and division, mostly through successive phosphorylation and dephosphorylation events that trigger into action certain biochemical growth factors and division mechanisms. In some instances, spatial redistribution of certain substances plays an important role. This is the case of fission yeast when protein kinase *pom1* is transferred towards the cell poles, which causes chain of chemical reactions leading to mitosis.

The second category of works focuses on more general issues, trying to find *conceptual* explanation and discover general mechanisms underlying growth and reproduction phenomena. Such approaches are discussed in reviews, Refs. 1, 6, and in articles Refs. 7-9. The central question of such studies is this: what is that fundamental mechanism that sequentially triggers growth and reproduction events, thus providing high stability and reproducibility of organisms' cell cycles? Most researchers support the view that size is relevant to division. The other school of thought emphasizes the importance of time interval between the birth of a new cell and the onset of its mitosis phase. However, despite numerous studies, the problem, which mechanism is responsible for such a *definitive* and *universal* behavior of growing and dividing cells, is not yet resolved. Both the "sizer" and "timer" models have their own drawbacks. In Ref. 8, the authors argue that the "sizer" model is substantially more stable and better corresponds to experimental data than the "timer" model.

On the other hand, the "sizer" hypothesis does not explain many known facts. The authors of review, Ref. 6, conclude: "Conversely, cell growth does not typically rely on cell cycle progression. In a great variety of cell types, when cell cycle events are blocked with chemicals or genetic lesions, growth continues unchecked." In other words, size alone does not guarantee definitive growth progression and triggering of division phase.

"Timers" and "sizers" concepts can be combined. For example, an experimentally well supported study[7] presents evidence that both cell size and cycle time show homeostasis.

There are several *fundamental* reasons why such single characteristics as size and time, and even their combination, cannot account for such a universal phenomena as growth and reproduction of cells. First, the growth and reproduction are *multifactor* phenomena, whose progression depends on *many* parameters, both internal and external ones. It would be reasonable to expect that evolutionary development resulted in more adaptive multifactor mechanisms responsible for growth and reproduction, than mere size and / or time. For instance, it is known that the mass of grown wild type fission yeast can differ by as much as four times depending on nutrient availability.[7, 9, 10] Strong dependence of cycle time on temperature is another factor that is taken into account by neither "sizer" nor "timer" models. Variability of duration of G1 growth phase, discussed in Refs. 6, 10, 11, depending on the original size and nutrient availability, is another effect that fits neither simple "sizer" nor "timer" models.

The second reason is the *universality* of growth and reproduction phenomena across many species and growth conditions. In this, we can take a look at inorganic world, which is governed by multitude of fundamental laws of nature acting at *different* scale levels. On the contrary, presently the main biological paradigm is that *all* properties of living species are defined by biomolecular mechanisms. This concept is in contradiction with the actual arrangement of things we observe in nature, which is governed by different laws acting at different scale levels.

The third reason is an extreme variability of forms and their geometric characteristics. We can see that size and shape of organisms can vary significantly in all three dimensions. Then, when it comes to size, its strict quantitative definition immediately faces big challenges and uncertainties; and this is not a technical issue, but a principal consideration. Even in case of relatively simple organisms such as *Schizosaccharomyces pombe* or *Escherichia coli* their lengths can vary significantly in the same growth conditions[7].

For similar reasons, "timer" models are unlikely to provide a credible answer.

So, if we want to find the fundamental cause of growth and reproduction puzzle, and through this to unlock the secret of life phenomenon in general, we should rather look for some general, of *fundamental nature*, mechanism (or mechanisms) that is universally applicable across different cells and unicellular organisms, and possibly tissues, organs, systems and whole organisms, and which inherently unites *combination of parameters* important for growth and division, which are necessarily present in *each* growth scenario.

Presently, overwhelming number of growth and reproduction studies are looking for solutions at a molecular level. This enthusiasm led to probably exaggeration of the actual role of biomolecular mechanisms in development of organic life and, on the other hand, to undervaluation of studies at other scale levels. Mitchison in Ref. 1 expressed similar concerns as follows: "Many biologists would agree that the shining gear wheels of the molecules have to be put back into the living cell to see how they really function there. The patterns of growth in volume and total protein or RNA provide an "envelope" that contains and perhaps restricts the gear wheels". We agree that parameters at different scale levels should interrelate; for instance, in our case, we found that compositions of biochemical reactions relate to distribution of nutrient influxes.

Ideas that there might be some mechanism influencing development of organic life at higher than molecular levels were presented in a classical work Ref. 12. The author said: "If the cell acts … as a whole, each part interacting of necessity with the rest, the same is certainly true of the entire multicellular organism". The author of Ref. 13 asserts: "… we are in danger of losing sight of an important fact: There is more to life than genes. That is, life operates within the rich texture of the physical universe". In our view, presently biology experiences certain lack of high level conceptual ideas, and it is about the right time to resume the search for explanation of growth and reproduction phenomena at other than biomolecular level.

In Ref. 1, the author sees the situation with the study of cell growth as follows: "It would be satisfying if the main parameters of cell cycle growth had been established in the earlier work. Not surprisingly, however, there were still major uncertainties left when people moved from this field to the reductionist approaches of molecular biology. This is particularly true of budding yeast and may raise problems for the survey methods of the future. In addition, the patterns of metabolism, e.g., oxygen consumption, are in a muddled state and very little is known of their controls in balanced growth - another subject for the future". Answering such valid concerns, here, we introduce some of such main parameters influencing the cell cycle, on one hand, and consider certain general mechanisms which define the cell cycle, using mostly growth and reproduction of fission yeast as an example, and to a lesser degree *amoeba*. In fact, we introduce rather a *framework* for studies of cells and unicellular organisms, illustrated by application to fission yeast *S. pombe* and *amoeba*.

The conceptual foundation of the study is a credible hypothesis, presently supported by many proofs, about the general growth mechanism, which we call the General growth law. This mechanism appears to be an influential player in life origin, development and evolution at scale levels from cells and cellular organelles to whole organisms. There is also evidence that this mechanism is also responsible for the balanced growth of different organs and systems in multicellular organisms.

We used experimental observations of fission yeast done by S. Baumgartner and I. M. Tolic-Norrelykke at Max Planck Institute of Molecular Cellular Biology and Genetics. Three data sets were provided, corresponding to different temperatures. In figures, the experiment number is denoted as "Exp. 1", "Exp. 2", etc.

## 2. Methods

### 2.1. *Conceptual foundations of the General growth law*

The General growth law was first introduced as a biophysical growth mechanism in Refs. 14-16 and in few earlier works. Following studies resulted in generalization of this discovery and introduction of the General growth law, which then was applied to different problems.[10, 17-21]

The main idea behind the General growth law may sound a little bit heretically for biologists: The biochemical machinery is not the only life creating and life supporting "tool", but rather it works in tight cooperation with the General growth law on this "assignment". Both work together, in inseparable *unity*, each at its level, doing its share of specific work. Their realms and functions do not intersect but *complement* each other. The General growth law is like an *upper* level management that provides general guidance for the biochemical machinery imposing certain *constraints*; for instance, specifying amount of nutrients that could be used for biomass synthesis. The existence of such a general mechanism removes the need for assumptions about existence of some genetic code controlling organisms' evolvement (presently a popular but never confirmed proposition), and explains known properties of growth and reproduction phenomena.

Mathematically, the General growth law is represented by the growth equation. Based on the General growth law, highly adequate and predictive growth models for unicellular organisms (*amoeba, S. pombe*, its mutant *wee1Δ*, *Saccharomyces cerevisiae*) were introduced in Refs. 10 and 21. Models showed very good correspondence with experimental observations, provided credible explanations of earlier observed phenomena, allowed predicting certain effects (which later found experimental confirmation) and discovering *actual* growth and division mechanisms. Its application to organs showed also high adequacy for modeling growth of livers and liver transplants for dogs and humans, and for finding their metabolic characteristics.[19, 20] These studies of organs' growth were also very important in that regard that they confirmed, through experimental and clinical observations, that the General growth law, indeed, acts at *different scale levels*, that is not only at the level of cells, but also at the level of organisms' organs and systems.

Based on the amount of accumulated evidence, we suggest that the General growth law is a genuine *fundamental law of Nature*, which (through imposing certain constraints biochemical mechanisms should comply with) much defines individual growth of living species and their constituents, and it is also an influential player in evolutionary development of organisms at different scale levels. The General growth law is based on two main properties of living organisms. First, in a growing organism and in its constituents and systems, the distribution of nutrients between two main biochemical activities, which are biomass synthesis and maintenance, is *uniquely* defined at each moment of growth and at each scale level. (Imagine an organism in which distribution of consumed nutrients between biomass synthesis and maintenance is arbitrary; such an organism would not survive and actually could not exist.) Such a unique distribution of nutrients provides their *optimal* use from the perspectives of fast growth for given conditions, adaptability, evolutionary development, individual growth, and it also secures *balanced* growth of different organs and systems within the same organism.

The next conceptual foundation of the General growth law is that the amount of produced biomass is a *leading* parameter of which composition of biochemical reactions strongly depends. There are many proofs supporting this statement. For instance, methods of metabolic flux analysis (analytical approaches widely used in biotechnological applications) produce the most adequate results when solution of system of stoichiometric equations is optimized for a *maximum amount* of produced biomass[10, 22]. This could be expected, since evolutionary pressures force natural selection in the direction of faster reproduction, which means faster growth and consequently the fastest possible rate of biomass synthesis. Of course, evolutionary optimization is always the result of compromise between different evolutionary constraints. However, for given constraints, the growth is optimized for a maximum production of biomass at *every phase* of growth. Inevitably decreasing amount of produced biomass during growth (for the reasons we will discuss below) secures *irreversible* change of compositions of biochemical reactions, through the growth and division cycles, and this is how organisms proceed through their life cycles.

Mathematical representation of the General growth law, the growth equation, uniquely defines this distribution of nutrients and unites it with biochemical and geometrical properties of growing organisms and their constituents through a set of parameters which necessarily present in each growth phenomenon (like acceleration, force and mass are the necessary parameters in Newton's second law and inherent to each mechanical phenomenon). From the physical perspective, one of the foundations of the General growth law is the law of conservation of mass, because nutrients are digested in biochemical reactions, for which the law of conservation of mass is valid.

The parameter, which defines distribution of nutrients between biomass production and maintenance, is called the *growth ratio*[10, 22]. It depends on the geometry of a growing organism and (indirectly) on its biochemical machinery. Numerically, it is equal to the fraction of nutrient influx that goes to biomass production. The value of the growth ratio monotonically decreases when organism's size increases. This property reflects on the fact that when biomass of an organism increases, more and more resources are required for *maintenance* of existing biomass, and accordingly a lesser fraction of resources is available for *biomass synthesis*. A certain critical minimum amount of nutrients (which depends on organism) that can be used for biomass production is what caps the grown size and eventually stops the growth.

Note that nutrients are always received through the *surface* (like cells' membranes, or surface of blood vessels), while they *always* support functioning of *volume*. When an organism grows, its volume increases faster than the surface area, even if an organism grows in one dimension, like fission yeast. This is why the nutrient supply per unit of volume eventually decreases during growth. Even if the nutrient supply per unit of surface increases (this happens in fission yeast and amoeba[10]), such an increase does not compensate for the volume increase endlessly, since no outside environment can provide unlimited nutrient influx.

The reduction of fraction of nutrients that goes to biomass production accordingly results in continuous *irreversible* changes in composition of biochemical reactions, since, as it was found in Ref. 17, this composition is tied to the amount of produced biomass. At some point, the fraction of nutrients used for biomass synthesis will reach a certain minimum value, to which a particular composition of biochemical reactions corresponds, and this triggers mitosis.

The General growth law directs forever ongoing biochemical activity in certain ways by imposing constraints. There is no purpose in these constraints; the General growth law is as purposeless as laws of classical mechanics are, or any other fundamental law of Nature is, for that sake, but these laws force things and phenomena to move on certain uniquely defined paths, like laws of mechanics force the thrown stone to fly on uniquely defined trajectory.

It is important for understanding the General growth law to know this. Cells and other organisms' components do not have separate sets of biochemical reactions - one specifically for maintenance and the other exclusively for biomass synthesis. Both activities are supported *by the same* biochemical machinery, in which all reactions interrelate [2,10,21]. Thus, any shift in nutrient distribution between these activities affects the *whole* biochemical machinery as a *single unity*. This fact is a key to understanding irreversibility of biochemical reactions during the growth cycle (and in many respects the life cycle too).

Diagram from Ref. 10 in Fig. 1 illustrates this arrangement graphically. Here, the growth ratio *uniquely* defines how nutrient influx is divided between maintenance needs and biomass production. Change of size and shape of an organism during growth causes change of the growth ratio through a feedback loop, which accordingly changes distribution of nutrients between biomass synthesis and maintenance. Composition of biochemical reactions is accordingly adjusted to the changing amount of produced biomass. Change of composition of biochemical reactions happens in phases. Each phase has some range of resiliency with regard to changing amount of produced biomass, while on the boundaries of such "resiliency ranges" composition of biochemical reactions changes more rapidly. Certainly, the "width" of such ranges differs in different growth phases. For instance, in division phase, biochemical transformations apparently proceed more rapidly than, let us say, in G1 phase. Eventually, when the amount of nutrients left for biomass production decreases to a certain value, this trigger transition to a new phase or stops the growth.

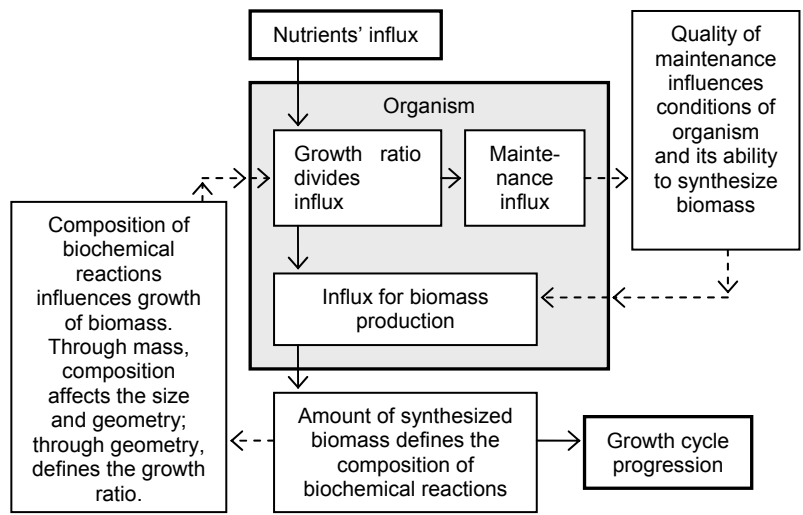

Fig. 1. How the General growth law manages growth. Growth cycle regulation and progression. (Copyright by the author, Ref. 10, Biophysical Reviews and Letters, Open access publication).

Although such an answer to the great puzzle of growth and life origin and development in general will be viewed by many people with skepticism, to say the least, presently the amount of available proofs presented in the aforementioned publications allows to suggest that the General growth law is one of that "secrets of life" people were aspiring to for so long. Unfortunately, it is not easy to understand and especially to comprehend its fundamentality as a life creating mechanism. However, if it was simple, it would be probably discovered long time ago. The truth is that unknown fundamental laws of nature *do exist*, and always will. Somebody may discover them from time to time, although such discoveries are rare; this is why people always have great difficulties recognizing them. Generality of fundamental laws of nature and rarity of their discoveries is what makes their introduction to the world *extremely* difficult, and in certain historical periods impossible.

## 2.2. The growth equation

The growth ratio parameter, which quantitatively describes distribution of nutrients between maintenance and biomass production, is defined as follows: Let us assume that nutrient availability and the biochemical specifics of a cell, which receives nutrients through its surface, allow it to grow to a maximum volume of $V_{MAX}$ with a maximum surface of $S_{MAX} = S(V_{MAX})$. (Generally, the maximum volume can vary, but we will address this issue later.) The dimensionless relative surface $R_S$ and the relative volume $R_V$ are defined as follows:

$$R_S = S(V)/S(V_{MAX}) \tag{1}$$

$$R_V = V/V_{MAX} \tag{2}$$

The dimensionless growth ratio $G_R$ is defined as:

$$G_R = \frac{R_S}{R_V} - 1 \tag{3}$$

Although the growth ratio is defined in geometrical terms, it is closely related to the biochemistry of cells, since it describes how much nutritional resources are used for growth (biomass production), and how much nutrients are used to support maintenance needs. A particular form of the growth equation depends on the growth scenario. When nutrients are directly supplied through the surface, then it can be written as follows[10,19,20]

$$p_c(X)dV(X,t) = \left( \int_{S(X)} k(X,t) \times dS(X) \right) \times \left( \frac{R_S}{R_V} - 1 \right) dt \tag{4}$$

Here, $X$ is the spatial coordinate, $p_c$ is the density of the tissue measured in $kg/m^3$, $t$ is time, $k$ is the specific influx, which is the specific nutrient influx per unit surface per unit time measured in $kg/(m^2 \times \sec)$, $p_c(X)dV(X,t)$ is the change in mass, and $dS(X)$ is an elementary surface area. In case when the specific influx does not depend on the location of an elementary surface area, equation (4) simplifies to

$$p_c(X)dV(X,t) = k(t) \times S \times \left( \frac{R_S}{R_V} - 1 \right) dt \tag{5}$$

where $S$ is the total surface through which nutrients are supplied.

Equation (4) has the following interpretation. The left part represents the mass increment. The right part is the total influx through the surface (the term $\int_{S(X)} k(X,t) \times dS(X)$), multiplied by the growth ratio $(R_S/R_V - 1)$, so that their product defines amount of nutrients that goes to biomass production.

As we said, the maximum size can vary depending on the growth conditions. Suppose that a cell grows in a nutritionally poor environment, so that it will have a small final size. If, during growth, the nutritional environment becomes richer, the cell's final size will be larger, which was experimentally proven in Ref. 11 and some other works. In many instances, the final size of a growing cell, or an organ, or an organism, can be known. For example, it can be found based on observations for particular nutrient content and concentration, or when some organ's mass is a known fraction of the whole body mass.[23,24] However, unless conditions for the whole growth period are known at the onset of growth, the maximum size is generally unknown. This fact does not mean deficiency of the growth equation, but reflects on how things are really arranged in Nature. The growth equation just describes this objective fact in a mathematical form. So, in a general case, the maximum volume is a function of other parameters, for instance, environmental factors, certain changes in organism's properties, like influence of physiological organ's function on the organ size, etc, that is $V_{MAX} = V_{MAX}(E)$, where $E$ is the combination of different parameters the maximum size depends upon.

## 2.3. *Modeling growth of fission yeast*

Long debated issue in fission yeast studies is what kind of mathematical function better describes the fission yeast growth - exponential, or linear, or bilinear or maybe some other.[25] Although the authors of Ref. 25, using their experimental observations and applying certain statistical data fitting procedures, favor a bilinear model (when the growth curve is approximated by two linear segments), they finally conclude: "growth of fission yeast does not follow a simple linear or exponential law. This growth is, instead, a rather complex process, the regulation of which will continue to intrigue researchers for some time to come".

In order to compute the growth curve for fission yeast using the growth equation (5), we need the specific nutrient influx $k$. It can be found through the total nutrient influx $K$ (defined by equation (A.12), derived in Appendix A), divided by the total surface $S$. We also need to know the growth ratio, surface and elementary volume. An adequate geometrical model of *S. pombe* is a cylinder with the beginning length $l_b$, ending length $l_e$, current length $l$, that has hemispherical caps at the ends with radius $r$. Unlike in equation (A.12), here, the lengths are measured in absolute values. In these notations, the relative surface, relative volume and growth ratio, defined by equations (1) - (3), are as follows.

$$R_S = \frac{(2r+l)}{(2r+l_e)}$$

$$R_V = \frac{((4/3)r+l)}{((4/3)r+l_e)}$$

$$G_R = \frac{R_S}{R_V} - 1 = \frac{((4/3)r+l_e)(2r+l)}{((4/3)r+l)(2r+l_e)} - 1 \tag{6}$$

If we introduce relative lengths' increases $L = l/l_b$ and $E = l_e/l_b$, and a relative radius's increase $R = r/l_b$, then equation (6) transforms to the following.

$$G_R = \frac{(2/3)R(E-L)}{((4/3)R+L)(2R+E)} \tag{7}$$

The volume is equal to $V = (4/3)\pi r^3 + \pi r^2 l$, so that the volume differential $dV = \pi r^2 dl$. If we use the relative length's increase of the cylinder $L$, then the specific nutrient influx $k$ (amount of nutrients per unit of surface) can be defined as in equation (A.12).

$$k = A(C_p L^2 + C_r L^3)/S \tag{8}$$

where $A$, $C$ are constants; index '$r$' corresponds to RNA, index '$p$' denotes proteins. If we denote nutrient fractions that go to RNA and protein synthesis accordingly as $C_r^s$ and $C_p^s$, then $C_p = C_p^s$, $C_r = (2/3)C_r^s$.

Note that equation (8) accounts for the cylindrical part of the model, but not for the whole length, which would include hemispheres. The reason is this. We may think of the cell's volume increase as due to the elongation of the *cylindrical* part, while hemispheres remain the same - they are just pushed sideways by the growing cylindrical part. So, the influx for the RNA and protein synthesis should be associated with the cylindrical part too. Substituting into the growth equation (5) the nutrient influx from equation (8), and the growth ratio from equation (7) produces the following.

$$p\pi r^2 l_b dL = A(C_p L^2 + C_r L^3)\frac{(2/3)R(E-L)}{((4/3)R+L)(2R+E)} dt \tag{9}$$

The analytical solution of equation (9) is presented in Appendix B. It is as follows.

$$t = \frac{B}{A}\left(f\left(\frac{1}{L_b} - \frac{1}{L}\right) + d\ln\left(\frac{L}{L_b}\right) + \frac{g}{C}\ln\left(\frac{1+CL}{1+CL_b}\right) + h\ln\left(\frac{E-L_b}{E-L}\right)\right) \tag{10}$$

where

$$B = 3p\pi R l_b^3 (2R + E)/(2C_p) \; ; \; f = \frac{4R}{3E} \; ; \; d = \frac{1 + f - fCE}{E} \; ;$$

$$h = \frac{d + fC}{CE + 1} \; ; \; g = C(h - d) \tag{11}$$

Although one would prefer to have a functional dependence of $L = L(t)$, the transcendental equation (10) cannot be solved analytically for *L*. However, it allows easily obtaining two sets of values for time and length using length as an independent parameter.

The set of input values which is required to compute the growth curve for fission yeast is presented in Table 1.

Table 1. Parameters required for computing the growth curve for *S. pombe*. Unit of length is $\mu m$. Values correspond to a particular experimental observation from Ref. 25.

| Parameters | Value |
| --- | --- |
| Initial length, $l_b$ | 10.2 |
| Grown length (corresponds to inflection point) | 18.25 |
| Maximum possible length, $l_e$ | 24.9 |
| Diameter, *2r* | 5 |
| $C_p$ | 0.714 |
| $C_r$ | 0.246 |

This is a *complete set* of parameters if an organism grows in a nutritionally stable environment. Otherwise, we have to know how the nutrients influx changes during growth and how this affects the maximum possible size. (Although here we assume that a diameter remains the same, if it is not so, the diameter's change should be also included into the growth equation.)

## 3. Results and discussion

In this section, we consider two distinct types of growth and division mechanisms, represented by *amoeba* and fission yeast *S. pombe*. Both mechanisms were discovered with the aid of the General growth law. Obtained results very well complies with knows experimental observations and explain known effects.

### 3.1. *Amoeba growth*

### 3.1. *Computing the growth curve*

Amoeba represents type of growth that uses probably the most basic and ancient growth and replication mechanisms, which include: (a) growth progression through almost the entire growth cycle predefined by the growth equation; (b) switching to division and progression through the division phase due to increasing rate of change of the growth ratio; (c) rates of synthesis of all cell components, such as proteins and RNAs, are the same.

Assumption about the same rate of RNAs and protein synthesis in *amoeba* is very likely to be true, given results of known studies analyzed in Refs. 8, 9 and 10 and results presented in Ref. 26.

Analysis of *amoeba* images and pertinent information done in Refs 10, 21 allowed to conclude that *amoeba's* growth can be well approximated by increase in two dimensions. So, *amoeba* was modeled by a disk whose radius increases while the height remains constant and equal to the initial disk diameter. A disk with a half smaller height was considered too, but the results of modeling were very close. It turned out that the most important factor influencing the shape of the growth curve is in how many dimensions the organism's geometry changes (for instance, a pinion-like shape produces almost identical results). For such a geometrical form and equal rates of RNAs and protein synthesis the nutrient influx is defined by equation (A.14) in Appendix A.

Using equations (1) - (3), we can find the relative surface, the relative volume and the growth ratio for the disk-like growing organism as follows.

$$R_S = \frac{r(r+H)}{R(R+H)}$$

$$R_V = \frac{r^2}{R^2}$$

$$G_R = \frac{R_S}{R_V} - 1 = \frac{R(r+H)}{r(R+H)} - 1 \tag{12}$$

Here, $r$ is the current radius of the disk, $R$ is the radius of the disk corresponding to the maximum possible volume, $H$ is the disk's height. For the growth equation, we also need to define volume's differential, which is $dV(r) = \pi H((r+dr)^2 - r^2)$, and the disk surface $S = 2\pi \times r(r+H)$.

Finding maximum possible volume for *amoeba* demonstrates some interesting growth features of such organisms. As a first approximation, one can use assumption that *amoeba* has to double its size. However, *amoebas*, experimentally studied in an excellent work Ref. 27, grew larger and smaller, although in one instance *amoeba*, indeed, doubled its weight (2.054 of the initial weight). We assume that *amoeba's* density is constant, which is supported by studies in Ref. 27, so that we may substitute mass for volume. It was found in Refs. 10, 21 that *amoebas* do not normally fully realize their growth potential, but divide slightly earlier. To account for this effect, the parameter "spare growth capacity" was introduced, which is interpreted as unrealized growth potential. It is defined as follows.

$$SGC = 1 - V_d / V_f \tag{13}$$

where $V_d$ is the volume when *amoeba* divides, and $V_f$ is the maximum possible asymptotic volume.

The value of spare growth capacity for amoeba is about 2%, ranging for experimental data from Ref. 27 from 1 to 2.8%. The spare growth capacity is not an abstract notion. Two *amoebas* failed to divide and, indeed, increased their weight by about 2% during the following growth. The reason that *amoeba* does not go through the whole possible growth cycle is that the growth in the final phase is very slow, actually asymptotic, so that it does not make evolutionary sense to wait for so long to gain so little weight.

Once we take into account the spare growth capacity of 2%, we can find the maximum possible volume from (13) as follows.

$$V_f = V_d / (1 - SGC) \tag{14}$$

Substituting nutrient influx defined by equation (A.14), the growth ratio from (12), and the maximum possible volume (14) into the growth equation (5), we obtain a differential equation, whose solution defines amoeba's growth curve. The solution can be obtained analytically or by solving the equation numerically (the approach we used).

From available experimental data in Ref. 27, we chose the growth scenario with stable nutrient supply[10,21]. (Of course, if we knew how the nutrient supply was changing, we would not have problem to adopt this information using the growth equation (5), but we do not have such data.) Fig. 2 presents the graph of a computed growth curve versus experimental data from Ref. 27. The only adjustable parameter was time scaling coefficient, which, of course, does not change the shape of the computed growth curve.

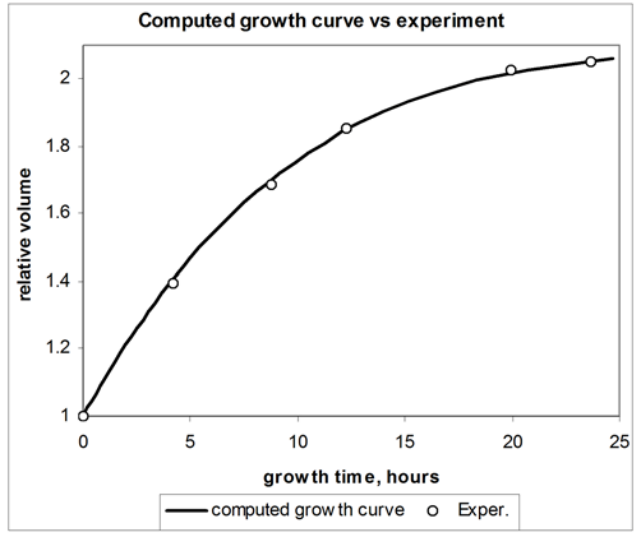

Fig. 2. Computed growth curve for *amoeba* versus experimental data from Ref. 27.

Note that *this is not a data fit procedure* in the usual sense. We *first* computed the growth curve, and only *after that* compared it to experimental data. Overall, we can observe good correspondence between the computed growth curve and the experiment, which confirms the adequacy of *amoeba's* growth model developed on the basis of the General growth law.

We also computed growth curves for other experimental observations, and found a reasonably good correspondence between experimental data and theoretical growth curves.

### 3.2. *Amoeba's division mechanism*

In this section, we explore *why* and *how amoeba* divides; what is that *fundamental* mechanism that forces division of organisms of this type. It was mentioned in Ref. 6, referring to Refs. 27-29, that it was possible "preventing Amoeba in G2 phase from entering mitosis indefinitely by periodically resecting a portion of its cytoplasm, thereby preventing the attainment of a presumed critical size". So, apparently, size matters for division, concluded the authors. However, so far all attempts explaining division by change of size did not succeed. Application of the General growth law to this problem provides the following answer. (It is definitely not the answer biologists would expect, but the convincing proofs are there, and it is a matter of overcoming the usual inertia of a human mindset in order biologists could accept this study and enter a new vast land of *tremendous* opportunities.) What matters for triggering the division phase is not exactly the size, but the *value of the growth ratio* which, indeed, depends on the current size and the maximum possible size.

The growth ratio defines the fraction of the total influx that goes to biomass synthesis. In turn, the composition of biochemical reactions depends on the amount of produced biomass. Table 2 shows the change of the growth ratio depending on the relative volume and time in the last four hours for the growth curve from Fig. 2.

Table 2. Absolute and relative change of growth ratio and volume depending on time.

| Time, hr | Volume | Volume change, % | Growth ratio | Gr. ratio change, % |
|---|---|---|---|---|
| 12.743 | 1.857 |  | 0.02186 |  |
| 14.192 | 1.8968 | 2.14 | 0.017304 | 20.84 |
| 16.074 | 1.9369 | 2.11 | 0.012843 | 25.78 |
| 18.758 | 1.9775 | 2.1 | 0.0084734 | 34.02 |
| 23.426 | 2.0186 | 2.08 | 0.0041934 | 50.51 |

For the data in the last two lines, which correspond to last 4.5 hours of amoeba's life cycle, we can see that the change of volume by 2% causes a reduction of the growth ratio by 50.5%, which is a significant amount. So, although the volume (mass) of *amoeba* changes by mere 2%, its growth ratio changes by *fifty* percent. Note that in the previous 2.7 hours the value of the growth ratio decreased by the substantial amount of 34% too. According to the General growth law, the amount of produced biomass is defined by the value of the growth ratio. So, significant relative changes of the growth ratio mean significant changes in the amount of produced biomass, since nutrient influx changes little at the end of growth cycle[10,21]. Amount of produced biomass, in turn, defines composition of biochemical reactions, as we previously discussed. So, drastic changes in the amount of produced biomass should accordingly mean significant changes in the composition of biochemical reactions, and this is exactly what happens in the division phase. This division mechanism explains the indefinite delay in amoeba's division by resection of part of its cytoplasm: *Amoeba* cannot divide unless its growth ratio (and accordingly amount of produced biomass, as a fraction of total nutrient influx) reaches a certain value, which corresponds to the composition of biochemical reactions required for the division, while resection was reducing the growth ratio below the threshold value required for triggering mitosis. Furthermore, this mechanism also explains *all* other known effects and facts related to mitosis and division of organisms which exercise the same type of division mechanism as *amoeba*. Here, we can also see that the composition of biochemical reactions corresponding to a certain amount of synthesized biomass (as a fraction of the total nutrient influx) corresponds to a certain *range* of growth process, a certain growth phase, within which some irreversibility in case of *amoeba* is possible.

Let us take a closer look. Although *amoeba*'s mass does not increase much at the end of the growth cycle, the *rate of change of the growth ratio* (which is equivalent to the rate of change of biomass synthesis) is *significant* until the end of the growth cycle. Even more, the rate of change *accelerates* towards the end of growth. This is an extremely important fact, because the growth ratio determines how much nutrients from the incoming flux are used for the *biomass production* $m_B$, as it follows from the growth equations (4) and (5). In order to see this in explicit form, we can rewrite the right part of the growth equation (5) as follows.

$$dm_B = k(X,t) \times S(X) \times \left(\frac{R_S}{R_V} - 1\right) dt \tag{15}$$

Is the substantial change in the growth ratio is that same factor that, indeed, governs the growth cycle and replication of *amoeba*? We stated a credible hypothesis that is well supported by available knowledge. Besides, in this regard, the General growth law, through the growth equation, reconciles *all* known facts.

Certainly, this hypothesis, despite its credibility and proofs, is still a hypothesis that can be transformed to theory only by its acceptance by a scientific community on the basis of many diverse *facts* and their interrelated *unity*. This is the only way to verify fundamental laws of nature, which are always *heuristic* discoveries, since no such prior knowledge existed. (Some physicists are still verifying the Newton's second law of mechanics.) However, the degree of novelty in case of the General growth law is a way too high in order for the recognition to happen soon (if ever), unless other people would join the author in his efforts to introduce the General growth law into the world. Today, people see it mostly as an interesting and quite adequate *model* (and models are what modern scientists used to deal with, most in their entire careers, nobody is to blame), but people do not see it as a *fundamental law of nature*, of which notion many people are unaware at all - unfortunately.

So, applying the General growth law to *amoeba*, we accomplished the following: (a) obtained an adequate growth model that very accurately describes experimental dependencies and explains *all* known facts about amoeba's life cycle; (b) suggested a credible hypothesis of *amoeba's* division mechanism, and, to some extent, confirmed it by known experimental observations.

### 3.3. *Computing growth curve for fission yeast. Division mechanism*

Analyses of 85 experimental growth dependencies for fission yeast from the study reported in Ref. 25, and doing their comparison with growth curves computed on the basis of the growth equation, we proved the earlier suggestion that the beginning of mitosis in fission yeast corresponds to inflection point of the growth curve. Two considerations support this result. First, computed fission yeast's growth curves and their comparison with experimental data show that triggering to mitosis, indeed, happens at inflection point of growth curves. Second, from the evolutionary perspective, the fastest possible growth (to which triggering mitosis at inflection point corresponds) provides the highest survival chances for the population (some conditions are applied, and in particular this is the ability of fission yeast to significantly decrease their size when nutrients are scarce. [13])

Let the DNA, proteins and RNA constitute 100%. Then, we may assume that DNA mass corresponds to 4%, and the 96% correspond to RNAs and proteins.[11] We did not find experimentally obtained values of protein and RNA contents for the fission yeast (another evidence of the overwhelming prevalence of a biomolecular paradigm, when biologists have been studying particular biochemical mechanisms, but paid no attention to integral biomolecular characteristics). To get an idea about the range of RNA content in unicellular organisms, we analyzed data for *E. coli* from Ref. 11, which shows the wide range (from 0.035 to 0.246) of RNA content. Of course, two organisms are quite different, but the only thing we needed in this case is that the range of RNA content, in principle, can be wide, which we confirmed.

However, for the growth equation, we need to know not RNA content, but the nutrient distribution between RNA and protein syntheses. Since the rate of RNA degradation differs from that of proteins for fission yeast[10,21], the fraction of nutrients that goes to RNA synthesis can be quite different from the RNAs' content.

Using the proposed approach, we computed a hypothetical *full growth curve*, as if fission yeast cell uses the whole possible growth cycle, predefined by the growth equation (although in normal conditions fission yeast divides earlier). The computed full growth curve and its first derivative (rate of growth), and their comparison to experimental data is illustrated by graphs in Fig. 3.

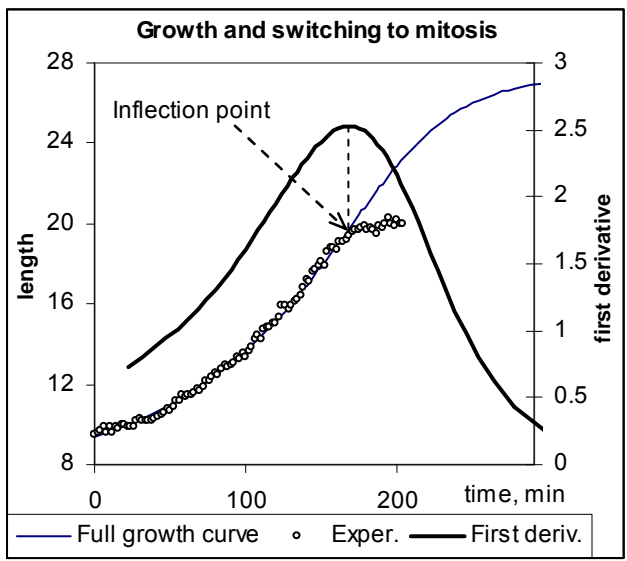

Fig. 3. Computed full growth curve and rate of growth (first derivative) for fission yeast versus experiments. Experimental data are from Ref. 25. Experiment No. 8, $32^oC$ data set. RNA nutrient fraction is 0.6; diameter is 4.6 $\mu m$. Length is measured in $\mu m$.

It follows from the graphs in Fig. 3 (as well as from similar computations for other 85 experimental datasets), that fission yeast normally switches to mitosis phase *prematurely*, at a *maximum rate of biomass synthesis* (which corresponds to inflection point of the full growth curve), without going through the whole possible growth cycle defined by the growth equation. Such a mechanism triggering mitosis phase is very different from the one in amoeba, which grows almost through the entire growth cycle predefined by the growth equation, as we have seen above.

We can see that at the limit the full growth curve goes to horizontal asymptote, which is the expression of the fact that, if fission yeast passes the point when mitosis begins, it will continue to grow, and so more nutrients will be required for maintenance of increasing biomass, and accordingly less and less nutrients will be available for biomass synthesis, which eventually will stop the growth. Indeed, when triggering to mitosis is blocked, fission yeast continues to grow further[6,25].

Note that as in case with *amoeba*, for fission yeast we also *first* computed the growth curve, and *only then* compared it with experimental data. So, this is not a data fit procedure in the usual sense, when some mathematical function is used for fitting, which was done in other studies, but we present the graph of solution of the growth equation for the fission yeast based on *real biophysical parameters*. The only adjustable parameter was time scaling coefficient in order to match units of time between the computed growth curve and experimental data. If we knew the nutrient consumption of fission yeast in absolute values, we would not need the time scaling coefficient.

Besides the fastest growth time achieved by switching to mitosis at inflection point, the well expressed inflection point effectively secures high sensitivity of the biochemical "mitosis trigger" in fission yeast discussed in Refs 3-5. Based on our study, we found that that the accelerated rate of RNA synthesis compared to proteins, and a cylindrical shape contribute to a well expressed inflection point. In fact, *amoeba's* growth curve has an inflection point too, but it is weakly expressed.

### 3.4. Blocking synthesis of DNA

The full growth curve presented in Fig. 3 is not a mathematical abstraction or an "ad hoc" model, but a *real* phenomenon. It was proved experimentally that when triggering to mitosis is blocked, fission yeast continues to grow *further*, beyond its usual grown size, or, if we refer to the full growth curve, above the inflection point. In Ref. 6, the authors acknowledged that "In a great variety of cell types, when cell cycle events are blocked with chemicals or genetic lesions, growth continues unchecked." This means that such cells exercise the same premature type of growth and division mechanisms as fission yeast does. Note that this type of overgrowth is not characteristic for *all* cells. Some organisms, like amoeba, will not exhibit such an overgrowth, since their growth cycle comprises almost the entire growth curve predefined by the growth equation. As we mentioned before, in an experiment presented in Ref. 27, two amoebas failed to divide and, indeed, increased their weight by about 2% *only*, which is consistent with the division mechanism we discovered on the basis of the General growth law.

In Ref. 25, the authors blocked DNA synthesis (S-phase). They found that "The HU-treated cells were, on average, significantly longer than untreated cells". Based on the results of the experiment, they say: "We conclude that the growth pattern of the cells blocked in S-phase consistent with linear growth."

These results agree with the growth curve computed for the suppressed S-phase on the basis of the General growth law. The computations were done as follows. It is reasonable to assume that, because only one copy of DNA is available in this case instead of two compared to normal growth, the rate of RNA synthesis reduces twice as well. Then, if we follow the derivation procedure similar to equation (A.14), the nutrient influx is as follows.

$$K(L) = AL^2 \left( C_p + C_r \right) \tag{16}$$

Substituting values from Table 1 and equation (16) into the growth equation (5), and solving it (using the approach presented in Appendix B), we find the following

$$t = \frac{B_S}{A_S}\left(f_S\left(\frac{1}{L_b} - \frac{1}{L}\right) + d_S \ln\left(\frac{L}{L_b}\right) + g_S \ln\left(\frac{E - L_b}{E - L}\right)\right) \quad (17)$$

where

$$B_S = 3p\pi\pi R_b^3 (2R + E)/(2(C_p + C_r)); \quad f_S = 4R/(3E);$$

$$g_S = d_S = (1 + f)/E; \quad (18)$$

Using equation (17), we obtain the growth curve presented in Fig. 4. We can see that, indeed, in the range analyzed by authors (from 10 to 100 min), the growth curve can be approximated by a straight line. We can also see that cells should grow noticeably longer.

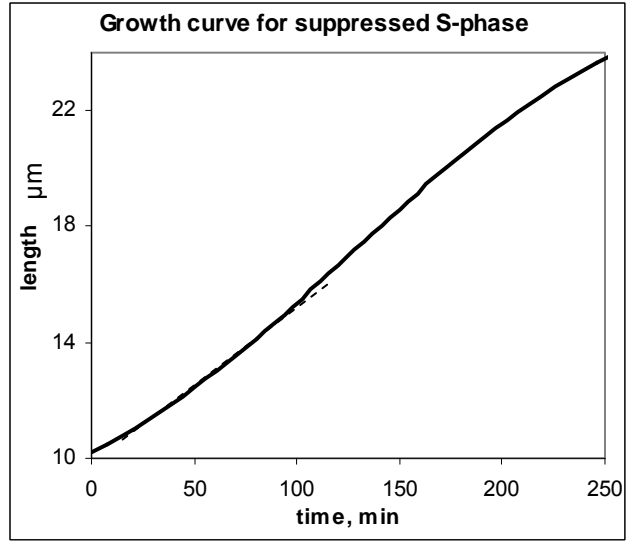

Fig. 4. Computed growth curve when S-phase (DNA synthesis) is suppressed. The dashed straight line closely matches the linearity of the growth curve observed in Ref. 25.

Thus, the General growth law, through the growth equation, explains the elongation effect experimentally observed in Ref. 25, as well as the linear character of the experimental growth pattern in the studied interval.

Since the normal growth cycle of fission yeast includes S-phase, the growth model of fission yeast (based on the General growth law) may include *both* curves. One is defined by equation (17), describing the part of growth curve before the RCP (rate of change point), which includes S-phase and apparently part of G2 phase, and equation (10), which describes fission yeast growth when DNA synthesis completes and both copies of DNA participate in transcription. The connection point of these two graphs is chosen at a time interval of 20-30% of the total growth period, depending on a particular growth scenario. According to experiments from Ref. 25, the length of S-phase can noticeably vary. The evaluations of the duration of S-phase are dissimilar in different studies[9,25,30], for which probably diverse growth conditions and different strains of cells could be the reasons. In any case, the range of 10 to 20% of the total growth period (before separation of new cells), and likely beyond, would include most estimations of time required for DNA synthesis. Results of previous studies allow to suggest that the end of S-phase does not necessarily coincide with the RCP, but could be considered as a prerequisite for the RCP to appear. For instance, in experimental growth curves presented in Ref. 10, the time intervals corresponding to RCP points are in the range of 32-34% of the overall cycle time, which is quite different from the results obtained in Refs. 25, 30, where the same intervals were mostly confined to 20%. Apparently, something else at a biomolecular level should happen before the rate of biomass production kicks higher gear. According to the growth equation, that "something" should relate to ability to process more nutrients, and to ability to synthesize more RNAs in particular.

### 3.5. *Statistical validation of the fission yeast growth model*

Based on experimental data, the geometrical form of the fission yeast' growth curve can be approximated by exponential function, by "bilinear" or "trilinear" models, presenting broken lines consisting accordingly from two and three segments. Analytical solutions of the growth equation, equations (10) and (17), representing *inverse* functions, consist of sums of logarithmic functions and hyperboles, so that the inverse functions should rather resemble exponential functions.

As we found from analysis of experimental data and growth curves computed on the basis of the General growth law, the bilinear character is more prominent for the fast growth scenarios due to high content of ribosomes and RNAs. This is also an empirical result obtained in Ref. 25. In the last study, the authors presented statistical analysis of different fission yeast's growth models and came to a conclusion that the bilinear model (when the growth curve is composed of two linear segments) provides better fit compared to other models (exponential, linear, bi-exponential), based on several criteria. The used criteria are all based on variance (an average sum of squared distances of lengths between the experimental data and assumed growth curve, for the same time). So, using variance for comparison of our model, based on the General growth law, and the bilinear one, is sufficient. In our case, the growth curve was computed using the growth equation. Authors in Ref. 25 used certain approximation functions.

Note that when one uses variance for model's *validation*, measurement errors should be taken into account. Such, even if some model objectively presents a real phenomena (let us say, the calculated trajectory of a thrown stone), but we measure time and stone's spatial location with large errors, then the variance, meaning divergence of measured trajectory from the computed, will be obviously large, but this fact does not mean that the laws of classical mechanics are invalid or valid approximately. So, the minimum requirement for making such a validation should be that the variance of experimental results from the computed growth curve should not exceed the measurement errors. Based on presented error bars in Ref. 25, and the character of experimental data, this condition in our case is fulfilled, that is the measurement errors are big enough.

Note that in our calculations of variance we were in very much disadvantageous position, in several respects, relative to the case of a bilinear model, namely: The bilinear model was *already optimized* for a minimum possible value of variance, while in our case we did time scaling (the procedure used for matching experimental data and the computed growth curve) as an average of time scaling coefficients, each calculated as the ratio of times corresponding to an experimental point and to the corresponding point on the computed growth curve (both points correspond to the same length). This led to a certain increase of ours values of variances, compared to a possible minimum value. Such, for all analyzed curves, the average value of displacement of algebraic sums of deviations (distances in $\mu m$ from experimental data to the growth curve, for the same time) was 0.0416 $\mu m$, while for obtaining optimal results it should be zero. However, we did not account for this disadvantage, and mention it just in order to confirm that our calculation of variance had no computational advantages compared to the approach used in Ref. 25.

Second, the authors of the aforementioned work used substantially smaller time interval for comparison. They say: "The interval $t$ = 10-100 min was chosen for fitting because the growth in the first 10 min was different and thus was treated separately. At $t$ > 100 min, nuclear division, and the accompanying slow-down of growth, was observed in individual cells, which changed the mean cell length of the population". In our case, we *always* considered the *whole* growth interval, until the mitosis begins and where the sharp turn of growth curves occur. Also, we start from 6 min instead of 10 min. (In fact, we would not have problem modeling growth from the very beginning, but the thing was that because of the discrete measurement time (2 min), we did not know when the actual separation happened.) Note that the actual growth time in our case *substantially* exceeded 100 min interval used for testing bilinear model. Such, for the group of measurements at $32^oC$, the average growth time was 140 min; for the group at $28^oC$ - 167 min, for $25^oC$ - 269 min. Besides, obviously non-linear character of experimental growth dependencies (which was especially pronounced for the experiments at $32^o C$) would significantly increase variance for the bilinear model, if the fit is performed for the full growth intervals, which we did. So, here, our approach is in great disadvantage too to the calculation of variance done for the bilinear model.

Third, for the bilinear model, the authors had no restrictions in choosing fitting coefficients disregarding the actual growth process, meaning dependence of fitting coefficients on real biophysical parameters, such as, for instance, nutrient influx, geometrical characteristics of fission yeast, growth time, rates of protein and RNA synthesis, etc. In our case, all these meaningful growth parameters were tied to each other through the growth equation, so that we could not arbitrarily change them. Furthermore, parameters have to be at least in the realistic range; some were fixed. For instance, we did not know diameters of cells, but we could not assume that the diameter, let us say, is 5 $\mu m$, when the fission yeast length at the beginning was 7 $\mu m$. Such a geometrical shape is not typical for the fission yeast. So, we chose the diameter about half of the initial length or slightly more based on photos of *S. pombe* from different sources. Also, we cannot assume that a substantial fraction of nutrients goes to RNA synthesis when the growth time is long, and vice versa. We could not adjust the nutrient influx, assuming it to be stable, although it could experience variations in reality, especially for long growth periods, and so on. In other words, we had very little freedom in choosing values of input parameters, compared to the bilinear model. Table 3 presents results of comparison of our model and the bilinear one.

Table 3. Comparison of variances between the bilinear model, Ref. 25, and the one developed on the basis of the General growth law. Numbers 32, 28, 25 denote temperature the measurements were made at; numbers in brackets denote the number of analyzed experimental curves.

|  | Bilinear model | GGL 32, subset (23) | GGL 32 full, (34) | GGL 28 (19) | GGL 25 (15) |
|---|---|---|---|---|---|
| Variance | 0.0248 | 0.0251 | 0.038 | 0.206 | 0.5219 |
| Time interval, min | 10-100 | 6-150 | 6-256 | 6-272 | 6-344 |
| Average time interval, min |  | 121 | 140 | 167 | 269 |

As we can see from Table 3, for the measurements done in the time interval of 6-150 *min*, our model produced almost identical variance value (0.0251 vs 0.0248), although the bilinear model was fit into smaller time interval (10-100 *min*). So, on this basis, given several comparison disadvantages, the General growth law describes growth of fission yeast better than the bilinear model. As far as bigger time intervals are concerned, for which we do not have bilinear model's data to compare to, the values of variance reflect on the value of measurement errors. For instance, the variance for the group GGL 25 is equal to 0.522, which is twenty times bigger than for the GGL 32 subset group. However, judging by appearance, the computed growth curves fit experimental data equally well in both groups, as it can be seen on the graphs presented in the Supplementary material available on the author's website. The difference in variances is due to bigger measurement errors in case of long periods of growth at lower temperatures.

So, given the fact that we did not have precise input data for our fission yeast growth model, and, at the same time, restrictive interconnectivity of these input parameters through the growth equation, which gave us little freedom in choosing their values, we would say that our model describes growth of fission yeast substantially better than the bilinear model (and according to Ref. 25 better than other models presented in that study). Besides, our model describes growth in the whole growth period, which in some instances was as long as 344 min, while the bilinear model was analyzed in a substantially smaller range of 10 - 100 *min*.

### 3.6. *Form of growth curve as a consequence of underlying biophysical and biochemical mechanisms*

One may argue that if we use some other approximation for the whole growth curve, let us say, an exponential curve $A\exp(bt)$, we might get better approximation than for the 10-100 min interval. We chose nine growth curves most suitable for the exponential regression by appearance from all three data sets, total

12. In each case, variances were *greater* than for the solution which the growth equation produced. However, we would like to pay the readers' attention to the fact that while our computation is tied to real parameters defining the fission yeast growth, the exponential data fitting has no relationship with the underlying real parameters the fission yeast growth depends upon. We may use a neural network approach and train the algorithm to fit most experimental points, so that the variance will be negligible. However, it is clear that computed this way, such a fitting curve will have no sense with regard to real growth phenomena. Even if some arbitrary regression fares better than the solution of the growth equation, by and large, it will be useless, since it will have no relationships with real parameters like length, diameter, nutrient influx, rate of RNA synthesis and so on. The nature of scientific quest is discovering the *core essence* of studied phenomenon and real, of core nature relationships, but not establishing the *likeness* of studied phenomena to other things.

The growth equation allows taking into account *actual* parameters the growth depends upon, so that in a well controlled environment the application of the General growth law could provide lots of opportunities for a *qualitatively* new understanding of fission yeast growth and division mechanisms and their interconnections. The General growth law and biochemical mechanisms work together, as a single unity. Combining both approaches could significantly accelerate and advance studies of fission yeast. In fact, the same consideration is valid for any other organism, because of the universal applicability of the General growth law to all organisms and its constituents.

Of course, many people will be skeptical reading these lines. However, this is the nature of fundamental laws of Nature, to which the General growth law belongs to, that they are always introduced into the world as *heuristic* discoveries, since no such prior knowledge existed, and so the judicial principle of presumption of innocence is not applicable to them - they are always assumed to be "guilty", unless it is proved otherwise.

### 3.7. *Some common features of growth and division mechanisms*

It is important to understand that the considerations presented above (as well as the whole material), are not speculations on the basis of some *model*, which by luck or coincidence explains known facts and allows discovering new phenomena. The General growth law and its mathematical representation the growth equation have nothing to do with models understood in the usual sense, that is as *approximations* of reality. In fact, these are fundamental laws of nature (and the General growth law is one of them), which create this reality, the world we live in. If not for this law, there would be no reality we know. So, when we speak about nutrient distribution, growth ratio, start of mitosis at inflection point etc, these considerations do have the same level of validity and fundamentality as notions of force, mass and acceleration in mechanical phenomena. The only difference is that notions of nutrient distribution between biomass synthesis and maintenance, growth ratio, leading role of biomass production in defining composition of biochemical reactions, etc. are less obvious than, let us say, acceleration, force and mass, whose actions, in one way or another, everybody experiences every day.

One comment during preliminary review of the article was about (biochemical) implementation of the General growth law. The whole chain of biochemical events that is forced into action by constraints on the amount of produced biomass imposed by this law is not known. For the fission yeast, thanks to works, Refs. 3-5, and others, we know *how* part of this machinery is working, but we do not know *why* it works this way. In order to answer this question, one should tie entire composition of biochemical reactions to the amount of produced biomass, similar to what metabolic flux analysis does. Then, one will see that decreasing amount of produced biomass at some point will lead to such a change of biochemical reactions that they will begin producing transportation agents, which will move *pom1* from the cortical nodes to polar ends. Of course, other factors can contribute to such a chemical change too, but their action in one way or another will be tied to the amount of produced biomass (more precisely, to the fraction of nutrients used for biomass production). This is the framework and the approach which will allow answering the question about biochemical implementation - for fission yeast, as well as for other organisms.

A premature triggering to mitosis at a maximum value of the growth rate is an evolutionarily newer and more advanced growth and reproduction mechanism. (Refs. 10, 21 present discussion of this interesting development). In case of fission yeast, the value of the growth ratio defines the amount of produced biomass and accordingly composition of biochemical reactions through the entire growth cycle, including division. However, the biochemical paths before and after the inflection point are different. When the growth stage approaches the inflection point, according to Refs. 3, 4, protein kinase *pom1* moves toward the poles, thus reducing its concentration at cortical nodes. This reduction of *pom1* concentration then leads to phosphorylation and inhibition of kinase *Wee1*, which accordingly removes inhibition of CDKs that start mitosis, so that they are activated and thus mitosis begins.

Biochemical reactions supporting cell cycle of fission yeast are strongly interrelated and, which is probably the most important consideration, chemically they are highly *reversible*, meaning that they can go both in forward and backward directions. One of the major biochemical mechanisms supporting this property are phosphorylation and dephosphorylation reactions[2,10]. On the other hand, life cycle progression in general, and of the cell cycle in particular, is fundamentally *irreversible*, and so the change of compositions of biochemical reactions is too. Otherwise, the progression through the growth phases and completion of division would be impossible, or at least impossible in such a definitive way, which we observe in nature. So, there should be an additional factor that secures this irreversibility, which acts like a ratchet prohibiting backward change of composition of biochemical reactions. Experimental observations show that the length increase, and consequently the mass increase, continues after the triggering of mitosis at inflection point of the growth curve (although the rate of growth significantly slows down). We argue that the continued *increase of biomass* is that mechanism that secures irreversibility of biochemical reactions and forces them to proceed through the entire division cycle. Note that the biomass increase after entering mitosis phase and until cellular division, though small, is fairly consistent across different experimental observations and is equal to about 2%, which is the same value that we found for *amoeba* - the fact that probably should be noted.

One may argue that irreversibility of the cell cycle can be independent of the cell growth. Indeed, there are cells that experience changes without growing, like early embryonic cells, as one commenter noted. However, these cells do not exist on their own but are constituents of an organism, whose other constituents grow, and so growth of other constituents may affect behavior of non-growing components, producing substances for "export" to non-growing constituents of an organism, thus triggering appropriate changes in them.

It is hard for many people to accept this view, since they are rather inclined to see biomolecular explanations of such phenomena. Consider the present view of cell cycle. In Ref. 31, the authors say: "Eukaryotic cells have evolved a complex network of regulatory proteins, known as the *cell-cycle control system*, that governs progression of regulatory proteins through the cell cycle. The core of this system is an ordered series of biochemical switches that initiate the main events of the cycle, including chromosome duplication and segregation." Then, the description of many interesting biochemical details follows. However, if we step on this path, then we would have to enter into an infinite loop of questions about what biochemical mechanism caused the previous one. And still, these will be the answers to questions "how", but not "why". The authors say: "Switches are generally *binary* (on/off) and launch events in a complete irreversible fashion". However, existence of these switches does not contradict the General growth law, which ties such triggering to the amount of produced biomass. Rather, it *enforces* this view. The subtlety is that irreversibility of these switches represents *biochemical implementations* of irreversibility of composition of biochemical reactions, enforced by the General growth law. On the other hand, if we drop the General growth law out of the picture and accept such a *simplistic* idea of sequential switches, then even in this case we are left without answering at least one question: what triggers the first of these switches? In a biomolecular realm, this can only be another biochemical switch, which is triggered by … another biochemical switch. It does not matter, how many such backward steps along the "switches" one will go, the beginning of such biochemical chain will never be reached. There are many reasons why there is no, and *fundamentally* cannot be, such a primordial biomolecular switch. Nature does not work this simple way, given the omnipresence, persistence (in very different conditions!) and enormous diversity of life forms. None of particular biochemical mechanisms could be so universal and iron clad stable in order to sustain enormous number of different factors and their combinations, random events and so consistently and precisely perform its duties to support growth and reproduction phenomena. The idea of one or a group of particular primordial biomolecular mechanisms is just the concept which is incommensurate with the grandeur and adaptability of life phenomenon as such.

Besides, in reality, the cell cycle is a complicated set of different chains of events and interactions of different mechanisms, which cannot be driven by simple sequential biomolecular events. The biochemical machinery has to be regulated as *a whole*, otherwise there is no way to synchronize and force cooperative working of so many mechanisms, which have to react to *enormous* diversity of outer and inner deterministic, as well as *random* events. The fact of life is that *there are* numerous chains of events and parallel working biochemical mechanisms in cells, each in its domain. It is impossible to assume that there is a controlling *biomolecular* mechanism, which could manage such enormously complicated activities. It should be something of *qualitatively* different nature acting at higher scale levels. On the other hand, the concept of the General growth law easily and naturally explains the growth and division phenomena at all scale levels, including fundamental mechanisms that control workings of biochemical machinery.

Nobody argues that *there are* sequential biomolecular events, for instance, age related, resulted in certain pattern of epigenetic methylation[32]. However, these are rather *consequences* than reasons, while the actual reason is unknown[a]. In this regard, the General growth law gives a very satisfactory answer, from the perspectives of many criteria, and not only for growth and division phases. In a grown multicellular organism, cells die and grow all the time, tissues, bones change, balance in nutrient consumption between organs and systems changes too. In other words, the base for action of the General growth law *remains* even in a grown organism. Of course, many more mechanisms, acting at different levels, support and direct existence of grown multicellular organisms. What we are trying to convey is that these are not biomolecular mechanisms alone.

Is the described growth and reproduction mechanism in fission yeast correct? We think so, because it explains *all* known facts about these phenomena not only in fission yeast but in unicellular organisms in general, and, which is *even more important* from the standpoint of theory of verification of scientific knowledge, seamlessly interrelates these facts to each other, so that the whole growth and reproduction phenomenon neatly sets within the framework provided by the General growth law. The fact of life is that biochemical machinery *does not* include instructions what to do next, and, if we think for a moment, it cannot. It supports relatively short chains of chemical reactions, similar to when alkaline and acid are mixed, but changing *composition* of biochemical reactions in order to secure progression through the entire life cycle is a task beyond such "chemical automation". For those who know how DNA is structured and functions, it is impossible to assume that this bulky and in many instances excessively large structure (amoeba's genome is about 200 times greater than in humans) contains something that could guide such well organized and persistent processes as growth and reproduction. Biochemical machinery can *react*, follow some *externally imposed* guidelines, but not more than that. Guidance has to be provided. In our view, this is what the General growth law does through imposing certain *constraints*.

The whole progression through the division cycle in *amoeba* is also due to *small* increase of biomass. However, the value of the growth ratio, which defines amount of produced biomass, changes *dramatically*, over 50%, during the same period, as we have seen. It is namely *this* drastic change of *relative* amount of produced biomass (relative to the total consumed nutrient influx) that secures the dramatic change of composition of biochemical reactions, which we observe during division phase. All these considerations allow concluding that, with very high probability, the division cycle in fission yeast is also driven by the change of the growth ratio, which accordingly changes amount of produced biomass. This change, in turn, causes successive irreversible transitions in composition of biochemical reactions during division phase.

---

[a] Actually, the mentioned study confirmed a hypothesis about ageing mechanism, which was inferred from several sources, including the General growth law. So that there is at least one reasonable explanation why such age related methylation occurs.

**3.8.** *Factors that contributed to development of an advanced growth mechanism*

According to the General growth law, at least three factors contributed to the evolutionary development of an advanced growth and division mechanism and a cylinder shape of organisms like fission yeast. The first is the accelerated rate of RNA synthesis (it exists already in *S. cerevisiae*, which, or similar organism, could be a possible predecessor of *S. pombe*). Analysis of the growth equation showed that the accelerated rate of RNA synthesis compared to the rate of protein synthesis makes inflection point of the growth curve better expressed, so that biochemical mechanisms can reliably "sense" it, and accordingly are less likely to miss it (which still may happen both in natural and artificially created conditions).

The next factor is a cylinder form. It was shown in Ref. 10 based on the General growth law, that *cylinder* like organisms have the shortest cycle time among all elongated forms whose initial length is about 1.6 of a diameter or more (for instance, like double cones or double frustums). Fig. 5 shows dependence of growth time for two initial lengths when an elongated form continuously transforms from a double cone to a cylinder. This is why evolutionary development quite logically led to eventual selection of a cylinder form - because it provides the fastest growth time.

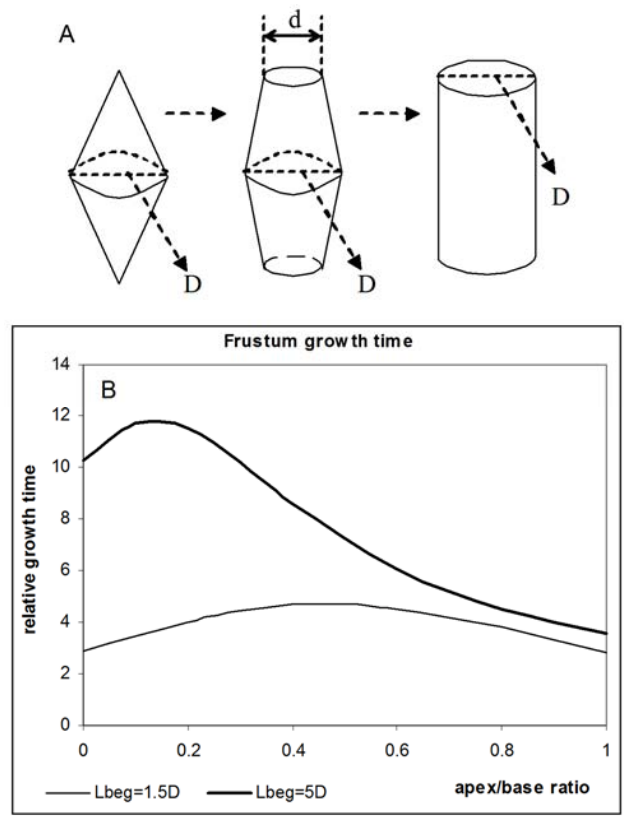

Fig. 5. Growth time of elongated form during transition from double cone to double frustum to cylinder. A - Change of geometrical form. B - Dependence of growth time on the ratio of apex to base diameter (d/D) for two initial lengths: 1.5D and 5D.

The third factor is the possibility to move protein kinase *pom1* from cortical nodes towards the polar ends, in order to reduce its concentration by about two times at the center, which is sufficient to trigger mitosis phase.

There are indications, which should be further studied, that it would be more difficult to do this in a sphere like organism.

### 3.9. *Length increase due to rounding of the new end*

Analysis of experimental data obtained in Ref. 25, as well as results presented in Ref. 9, showed that within first several minutes *S. pombe* experiences fast length increase. In Ref. 10, it was incorrectly assumed that this effect was due to certain specific nutrient requirements for synthesis of DNA. In fact, the effect is almost surely due to rounding of the new cell end, which is pretty much flat at the moment of cell separation, and then rapidly "inflates" to a hemisphere shape. This phenomenon was experimentally and theoretically studied in Ref. 25. So, the initial growth phase includes both length increase due to this rounding of a new end, and also the overall growth of an organism, which, according to Ref. 11, starts immediately too. The experimentally observed in Ref. 25 increase was about 7%. If we take into account the overall growth, then, according to our calculations, the input of the rounding of the new cell's end to the total length is in the range of 3-6%. The difference is most likely due to discretization of measurements (2 min intervals), so that some cells already increased their length due to the rounding of a new end, before the measurement took place. Presented results of comparisons of theoretical growth curves and experimental data often show such a rapid initial increase of *S. pombe* length, so that this effect should not be treated as the model inadequacy. It is easy to include this length increase into the model, but the problem is that we do not know when the actual cell separation happened, because of the discretization of length measurement. Fig. 6 shows a computed growth curve for experimental data from Ref. 9 for the wild type fission yeast. We can see exactly the same effect of faster length increase at the very beginning of growth due to rounding up of the new end. We also can observe good correspondence between the computed growth curve and experiment.

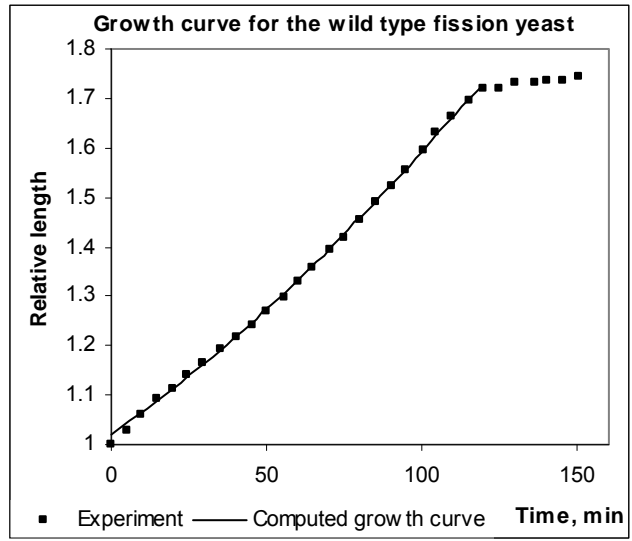

Fig. 6. Growth curve for wild type fission yeast versus experimental data from Ref. 9. $C_p^s = 0.824$; $C_r^s = 0.136$.

Time of such length increase is greater for lower temperatures: at temperatures of about $25^0 C$ it takes up to 10-12 min; at $28^0 C$ - up to 6-7 min; at $32^0 C$ - about 4 min. If such an increase is to be taken into account in the growth model, then the dependence should include temperature as a parameter and the cellular division time has to be measured more accurately, probably with an error of not more than 10-15 sec.

### 3.10. *Influence of diameter*

Experimental observations on growth and reproduction of fission yeast do not usually include both diameter and length for the same organism. The reason is that it is assumed that during growth the diameter does not change or changes little, from which the inference was made that knowing diameter of fission yeast is of no great importance. Technical difficulties certainly contributed to the problem, but, in our opinion, the idea that measuring diameter is of no great importance, also discouraged researchers to look for a solution harder than they did so far. In fact, two organisms with the same length but different diameters have noticeably different growth curves, since such organisms have different ratios of relative surfaces and relative volumes, which, according to the growth equation, results in different growth ratios and consequently in different rates of growth through the whole growth period.

Fig. 7 shows such growth curves computed for two hypothetical organisms modeled by cylinders with hemispheres at ends that have the same overall lengths, but different diameters of 5 $\mu m$ and 4 $\mu m$ respectively. As we can see, their growth curves are noticeably different. So, measuring diameter of fission yeast during studies, besides its length, is important for adequate and unambiguous modeling of fission yeast growth and other studies. Note that the graph corresponding to an organism with smaller diameter has lesser curvature.

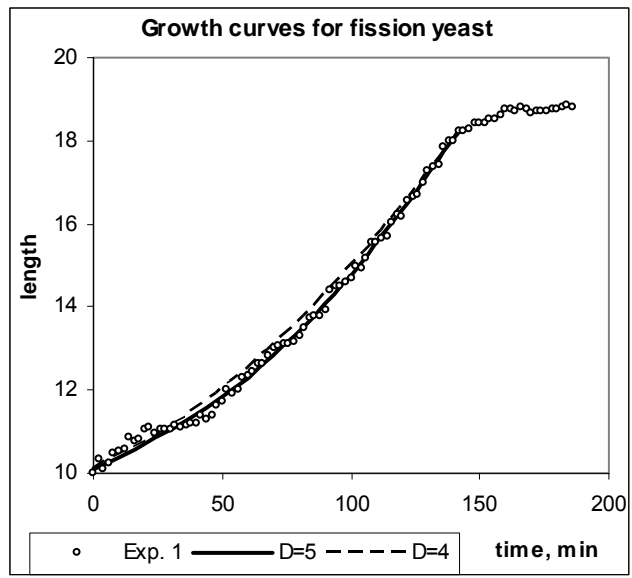

Fig. 7. Computed growth curves for two values of diameter. Diameters are equal to 4 $\mu m$ and 5 $\mu m$. Experimental data from Ref. 25.

### 3.11. *Finding amount of nutrients for RNA production*

The role of ribosomes and RNAs in cells is important for protein synthesis. It was shown in Ref. 11 that big and fast growing *Salmonella typhimurium* cells have substantially higher ribosome content than the small slow growing cells. Referring to photos, the authors say: "Note that the cytoplasm of the "fat" cells (a) from the rapidly growing culture is uniformly packed with ribosomes, whereas in the "skinny" cells (b) from the slow growing culture the few ribosomes present tend to cluster around the nuclear regains." Although we consider a different organism, the generalization of the known experimental facts that bigger organisms of the same type generally have greater RNA content than smaller ones, can hardly be argued against, since bigger mass means higher biomass production rate, to which RNAs are contributors. Higher temperature is also a factor contributing to elevated rate of RNA production. Due to the differences in rates of synthesis of proteins and RNAs, it is possible to find nutrients consumed for protein and RNA synthesis separately using the General growth law, through the growth equations (4) or (5). Below, when doing such computations, we assume that the density is equal to 1 $g/cm^3 =$ 1 $pg/\mu m^3$, where 1 $pg = 10^{-12} g$, which is consistent with experimental observations that the cell's density is close to density of water, for instance, in experimental studies[27]. Fig. 8 shows computed hypothetical growth curves for growth scenarios of fission yeast, which differ by a fraction of nutrients that go to RNA synthesis. This is the case of slow growth, so that the amount of nutrients that are used for RNA synthesis should be small, according to Ref. 11. Indeed, as we can see from the graphs, the most adequate curve corresponds to the lowest value, that is to 0.02 (a fraction of nutrient influx used for RNAs production).

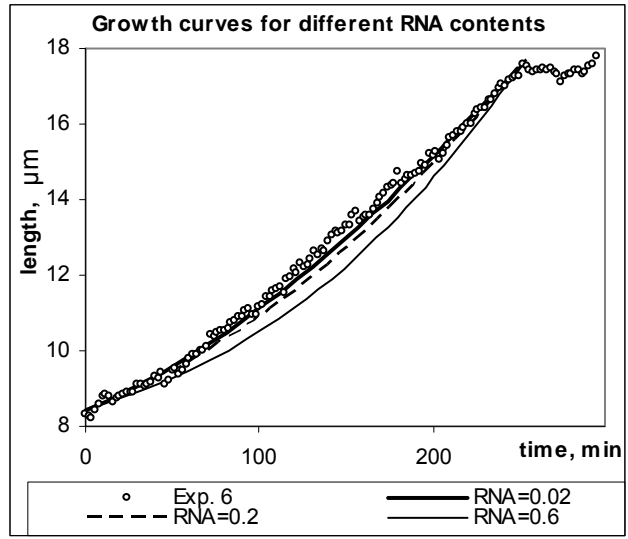

Fig. 8. Growth curves for different values of nutrient influx that go to RNA production. Experiment 6, $25\ ^0C$ data set from experimental studies in Ref. 25.

Since we did not know the diameters of analyzed organisms but can only make certain guesses based on the their length, we could not accurately estimate nutrient influx that goes to RNA synthesis. So, we will talk rather about tendencies than the actual numbers. The rate of RNA synthesis correlates with the cycle time. The cycle time, in turn, correlates with temperature. Nonetheless, there were two slow growth scenarios in the $32\ ^0C$ data set (total growth times 288 and 286 min, while the average growth period is about 140 min). For them, the computed RNA fraction was about 0.04. (Maybe this low value indicates that the data were included in the wrong dataset?) About 51% of experimental data within the same $32\ ^0C$ dataset were well fit when the fraction of nutrients going to RNA synthesis was in the range 0.6-0.7. About 20% corresponded to the range 0.2-0.3, for the rest the result was uncertain.

In case of $28\ ^0C$ data sets, which actually united range of temperatures, Ref. 25, in 55% of observations the RNA fraction of nutrients was in the range 0.1-0.3, and for 35% in the range of 0.02-0.04. In the $25\ ^0C$ data sets, about 58% of experimental data were well fit when the RNA's fraction of nutrient influx was only 0.02-0.03, for the rest no calculations were performed because of the "concave" character of data (which we will consider later) and other experimental data inconsistencies.

In all cases, the value of the nutrient's fraction for RNA synthesis correlated with the growth rate, meaning a percentage of length increase in a unit of time. Many cells growing at $32^0$ C did not increase their length twice, but divided noticeably earlier.

One of the inferences of our studies is this. The value of nutrient fraction that goes to RNA synthesis apparently does not influence the grown size much. If the initial size is small, organisms can still grow to more than double the initial size even if this fraction of nutrients consumed by RNA synthesis is relatively low (in the range of 0.02-0.04). Another observation was that there is a positive correlation between the small fraction of nutrients that go to RNA production, and the large cycle time, which is understandable, given the role of RNA in synthesis of biomass.

The question may arise: why the amount of nutrients that go to RNAs synthesis can differ by as much as 10-20 times, while the appropriate change in the overall cycle time is only about two times? This can be explained as follows. The degradation rates of RNA strongly depend on temperature, which in turn defines the growth rate. In Ref. 11, the authors cite results of several works in this regard. In particular, they acknowledge: "When measured at different temperatures the rate of decay was found to vary approximately as the growth rate, or the rate of protein formation. Thus, the "messenger" half-life was determined to be about 0.7 min at $37^0C$, about 2.5 minutes at $25^0C$, and about 1 hour at $10^0C$", referring to work by Fan, Higa and Levinthal done in 1964. In other words, the turnover rate of mRNA greatly increases with the rise of temperature, which accordingly requires substantially more nutrients for mRNA synthesis, and most likely for other RNAs too. It was also found that the decrease in energy metabolism "greatly increases" the stability of the total messenger as well as specific messengers[11]. So, the substantially stronger stability of mRNA, and very likely other types of RNAs, at lower temperatures, apparently compensates for the substantially less amount of nutrients required for RNAs synthesis in case of slow growth scenarios, and accordingly explains the sharp increase in the amount of nutrients required for RNAs synthesis when temperature increases.

The small fraction of nutrients that go to RNA synthesis partly explains what causes the bi-linear character of fission yeast's growth curves to be less expressed at lower temperatures, which was noted in Ref. 25. This happens because at lower temperatures, as we found, the fraction of nutrients that go to RNA production is small, and accordingly the value of the second term with cubic dependence on relative length's increase in equation (A.12) is small as well. On the other hand, it is namely this term that causes a sharp increase in the rate of growth when the RNA's nutrient fraction is high, which accordingly causes a better expressed bilinear character of growth in case of higher temperatures. Example of a growth curve when RNA's nutrient fraction is low is shown in Fig. 9.

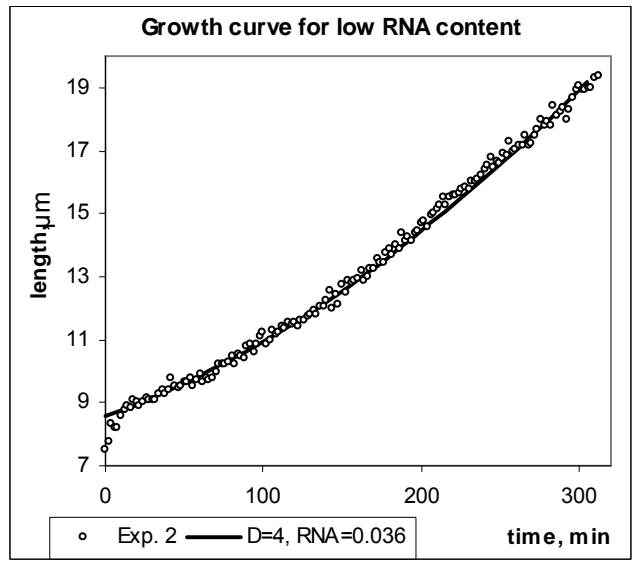

Fig. 9. An example of growth curve versus experiments. Slow growth and low value of RNA's nutrient fraction. Data from Ref. 25. Experiment 2, $28^0$ C dataset. Diameter $D = 4 \mu m$. RNA fraction of nutrients is 0.036.

Overall, the proposed method for evaluation of how much nutrients are consumed by RNA synthesis showed its practical value. For accurate evaluation, knowledge of cell's diameter is required. In many instances, the diversions of length measurements in Ref. 25 were too large for accurate determination of nutrient influx that goes to RNA synthesis.

Thus, so far, we see that the growth is defined by length and diameter, and by biochemical characteristics, such as how nutrients are distributed between the RNA and protein synthesis, which, in turn, depends on temperature and organism's specifics.

**3.12. *Metabolic properties of fission yeast***

One of the important methods composing this framework relates to cells' metabolism, which we illustrate by the study of fission yeast's metabolic properties. In general, knowledge of integral metabolic characteristics of organisms is very important, since this allows interconnecting more specific mechanisms and parameters through such "common denominators". In turn, this provides a solid foundation for different mechanisms, which can be used for reconstructing the whole structure of an organism, its morphological parts and functional blocks. The General growth law could one of such important foundational blocks.

Metabolism of organisms and their constituents relate not only to lower level metabolic mechanisms, but also to integral characteristics, such as organism's size. For instance, it is well known from allometric studies that organism's metabolic rate depends on size[33-35]. In turn, metabolic and other properties of organisms play an important role in evolutionary development of species, their parts, organs, systems, etc.[34,36] Knowledge of how to find certain integral characteristics, and especially how they interrelate to each other would provide many insights to the foundational mechanisms governing growth, reproduction, development and evolution of living organisms and their constituents.

On the basis of the General growth mechanism, we propose a method which interrelates geometrical and growth characteristics of cells and their integral metabolic characteristics, and exemplify the use of the method by applying it to fission yeast. Using the same General growth mechanism, we consider relationship of growth and geometrical form with evolutionarily acquired traits in fission yeast.

The General growth law allows evaluating both differential and total nutrient consumption by living organisms and their constituents. What is probably more important, it allows separately finding out how much nutrients go to biomass production and how much nutrients are used for maintenance. It is possible to find the amount of produced biomass and accordingly consumed nutrients based on the known mass increase. However, it is not easy to answer the question of how much nutrients are used for maintenance. On the other hand, metabolic consumption for maintenance is a very important characteristic, both from practical and theoretical perspectives. In this regard, the General growth law, indeed, provides a unique opportunity.

Previously, a method, which is also based on the General growth law, was suggested for finding metabolic characteristics of *S. cerevisiae*[17]. Later, similar idea was exploited for finding metabolic characteristics of dog and human livers[20]. For that, the model developed for liver and liver transplants growth was used, introduced in Ref. 19. The approach produced credible and confirmed by experiments and clinical observations results.

Using the growth equation (5), we can find nutrient influxes per unit of volume required for maintenance ($K_m$) and biomass synthesis ($K_{gr}$) as follows.

$$K_m(t) = k(t)S(t)(1-G_R(t))/V(t) \tag{19}$$

$$K_{gr}(t) = k(t)S(t)G_R(t)/V(t) \tag{20}$$

Here, $G_R(t)$ is the growth ratio as a function of time *t*.

Cumulative amounts of nutrients for maintenance *M* and for biomass synthesis *B* for a period $[t_1, t]$ are as follows

$$M(t) = \int_{t_1}^{t} V(\tau)K_m(\tau)d\tau \tag{21}$$

$$B(t) = \int_{t_1}^{t} V(\tau)K_{gr}(\tau)d\tau \tag{22}$$

Fig. 10a shows dependencies of influxes per unit of surface and per unit of volume. As we can see, during growth, nutrient influxes per unit of surface and per unit of volume increase, which means that during growth metabolic processing capacity of fission yeast per unit of volume substantially increases (about 4-6 times, depending on the growth scenario).

Fig. 10b shows the change of nutrient influxes consumed by fission yeast separately for growth and maintenance, and the total influx, depending on time. Towards the end of growth, nutrient influx for biomass production tends to go to horizontal asymptote, while the rate of increase of maintenance influx continues to grow. We can also see that substantially larger fraction of nutrient influx is used for maintenance needs than for biomass synthesis.

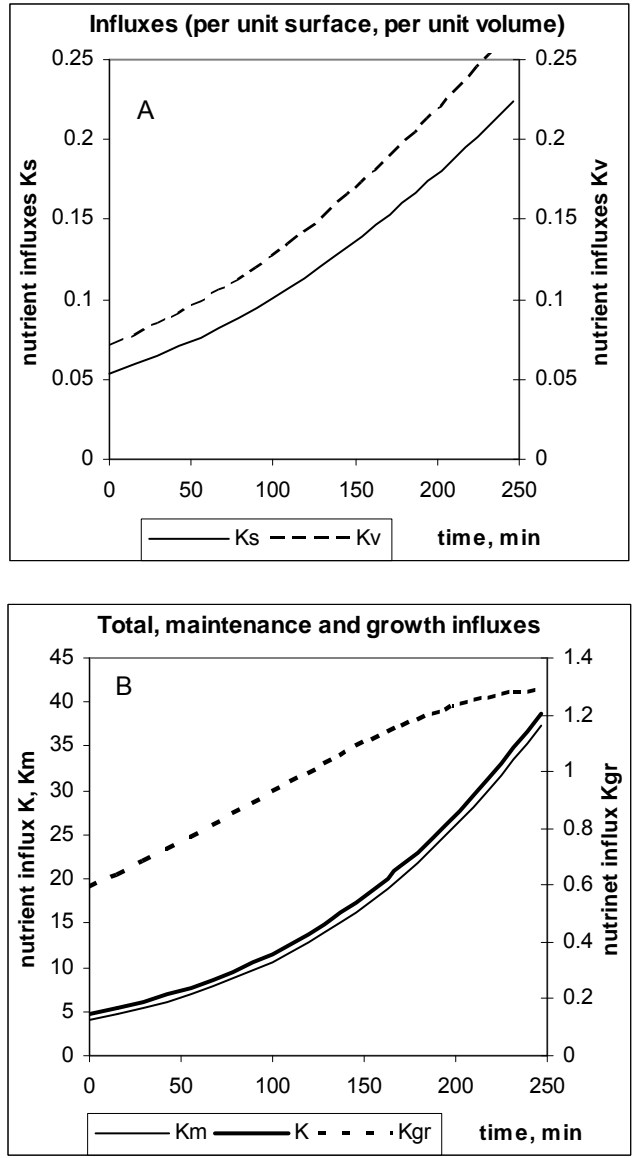

Fig. 10. Dependence of nutrient influxes on time. A - influx per unit of surface $K_S$ $(pg/(\mu m^2 \times \min))$; influx per unit of volume $K_V$ $(pg/(\mu m^3 \times \min))$. B - influx for biomass production (growth) $K_{gr}$, maintenance influx $K_m$, total nutrient influx $K$, all measured in $(pg/\min)$. Experiment No. 21 from $25^0 C$ data set[25].

Fig. 11 shows accumulated amount of nutrients used by fission yeast for biomass production and maintenance during the whole growth cycle.

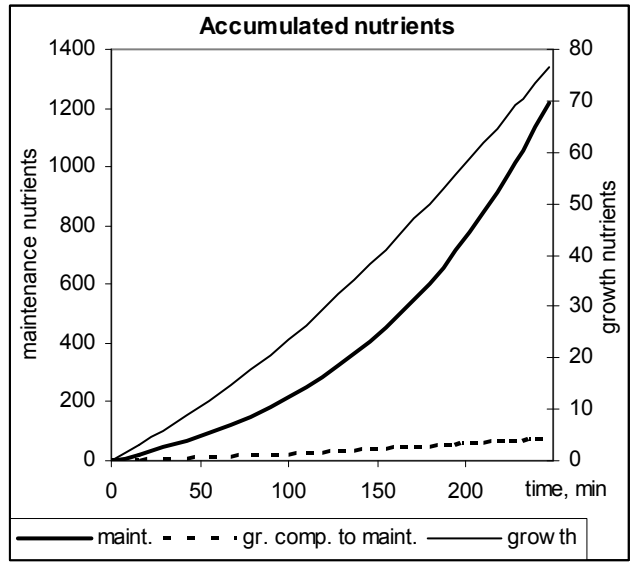

Fig. 11. Total amount of consumed nutrients for growth and maintenance. Nutrients are measured in $pg$. Experiment No. 21 from $25\ ^0C$ data set, Ref. 25.

Nutrient consumption for biomass synthesis calculated on the basis of the growth equation was about 76.34 $pg$. Obviously, this amount has to be equal to the overall mass increase found *independently* from direct experimental observations (through the increase of volume, for the density of 1 $pg/\mu m^3$). This equality, indeed (!), is fulfilled. This is an important validation step of the proposed approach. In fact, it confirms not only the validity of the proposed method, but also the validity of the growth equation (5) and its analytical solutions (10) and (17) with all preceding considerations; in other words, the validity of the General growth law and the growth equation.

Note that the amount of nutrients that goes to maintenance is significantly greater, about 1219 $pg$ versus 76.34 $pg$ that is used for biomass production, roughly by 16 times more. Knowledge of these parameters allows to accurately evaluate total amount of nutrients required for producing certain amount of biomass, which is important in biotechnological and other applications. This approach can be beneficially combined with a new advanced method for modeling population dynamics (meaning both population growth and reduction), which takes into account specific growth and metabolic characteristics of *individual* species and uniquely translates them into properties of *populations*[18].

Note that the more the relative increase in length during growth, the greater is the relative difference between nutrient consumption for maintenance and biomass synthesis. For instance, for the $25\ ^0C$ experimental data, this value can be as large as 18.1 times. This is because a greater fraction of nutrients is required to cover higher transportation costs in longer cells, due to longer distances the substances are transported to (see Appendix A). On the other hand, rate of growth and proportional increase in size does not influence cumulative distribution of nutrients between maintenance and biomass production.

### 3.13. *Concave growth curves*

There are several experiments in Ref. 25 when data sets are better approximated by *concave* curves rather than by the typical *convex* ones. There are 2 such experimental observations in the $32\ ^0C$ data set out of 41, one in $28\ ^0C$ data set out of 20, and 5 in $25\ ^0C$ data set out of 24 measurements.

The reason that the growth curve can have such or even more exotic shape is most likely the change of rate of biomass production, if we discard technical observational issues. According to the growth equations (4) and (5), it can occur because of the change of the growth ratio, or because of the change of nutrient influx (in case of a concave shape it has to decrease), or both. Change of growth ratio in case of fission yeast requires substantial changes in geometry, which are unlikely (fission yeast consistently preserved cylinder shapes with hemispheres at ends in experiments, according to Ref. 25), or substantial changes in biochemistry of organisms, like some malfunctioning. So, the change of the growth ratio, although theoretically possible, could not probably occur. Thus, the likely and, in fact, quite possible reason could be the change of nutrient influx.

We modeled such a situation using the following approach to describe the decrease of nutrient influx.

$$K(V) = K_0 (2.4 - 2.3(V - V_b)/(V_e - V_b))^3 \qquad (23)$$

Here, $K_0$ is a constant. Such a functional dependence describes an "upside-down" cubic parabola located above the abscissa. The appropriate "concave" theoretical growth curve is shown in Fig. 12. It closely mimics actual experimental dependencies.

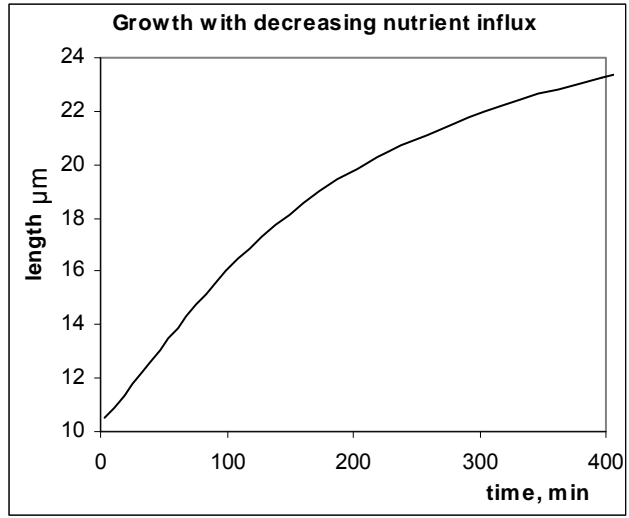

Fig. 12. A theoretically computed "concave" growth curve. Diameter is 5 $\mu m$, fraction of nutrients that goes to RNA synthesis is 0.36.

So, indeed, a concave shape of the growth curve is possible in certain conditions, when nutrient content acquired by an organism decreases by several times during growth. This may happen either because of the decrease of nutrients in the surrounding environment, or because of inability of an organism to acquire and properly digest nutrients, let us say because of some biochemical malfunctioning, which could be induced or naturally occur.

### 3.14. *S-phase. Specifics of modeling*

Not all growth curves correspond so well to experimental data as in Fig. 3. At high temperatures, the fraction of experimental data before the RCP (rate of change point) in some instances, especially at high growth rates, has slightly smaller slope than the theoretical growth curve predicts, which is illustrated by Fig. 13.

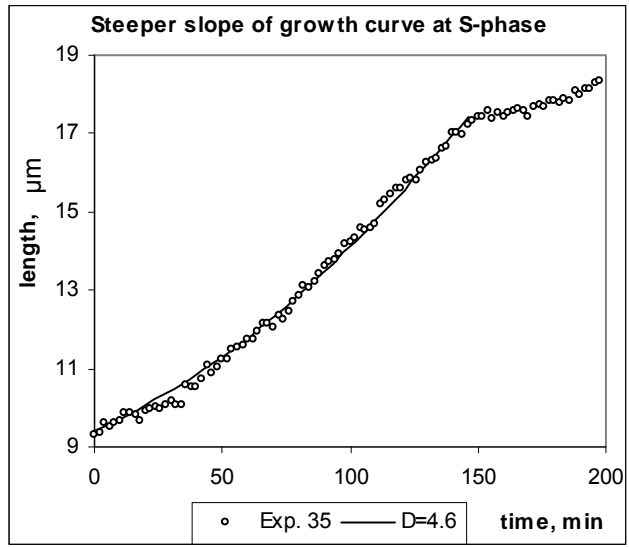

Fig. 13. Growth curve predicts steeper slope at the beginning than experimental data show. Experimental data from Ref. 25. Experiment 35 from $32\ ^0C$ data set. Nutrient consumption for RNA synthesis is 0.2.

One of the reasons of such behavior could be changes in geometry, namely increase of diameter when cells grow bigger, which would produce steeper slope after RCP. This seems as a reasonable assumption, given the fact that, depending on growth conditions, the same strains of fission yeast, indeed, may have different diameters in different generations. To clarify this issue, measuring diameter is needed.

The experimental fact discovered in Ref. 25 that temperature does not influence much the shape of the growth curve at S-phase may relate to the strictly ordered and complex process of DNA synthesis, in which temperature sensitive mechanisms do not play much role. This is the case when, evolutionarily, the accuracy of copying would be more important than the speed. Still, this is an issue to be studied.

**3.15.** *Evolutionary paths of microorganisms from the perspective of General growth law*

One of the interesting areas of application of the General growth law and the growth equation are evolutionary and development studies. Here, we demonstrate this application by comparing evolutionary paths of fission yeast *S. pombe* and *S. cerevisiae* from the perspective of the General growth law. Previously, the idea was discussed in Ref. 10.

Evolutionarily, in order to provide fast growth, first organisms had to originate on the basis of the most fundamental growth and division mechanisms. *Amoeba* apparently has such ones. This organism entirely relies on the most foundational mechanism, which is the change of the amount of produced biomass. One may argue that *amoeba*, as eukaryote, appeared far later than the first prokaryote originated, and so it should not have the basic division mechanism. Well, why not if such a mechanism works well enough to secure reproduction of so many species for so long. Evolution of fundamental level growth and division mechanisms is not a linear process in time. Certain *quantitative* changes may accumulate for a long period before their entire influence could produce a new *quality*. With regard to these new qualitative states, evolution is a discrete process, and the discretization interval can be very big. A bow and arrows are still in use, as well as a hammer. Animals use teeth and claws for fighting and hunting for a very long time, since these "tools" work well for them. It is not the age of a mechanism or a tool that defines its usage, but efficiency. So, there is nothing wrong that *amoeba* might have the most ancient division mechanism, if it worked so well for it and other organisms for so long.

Based on the study of the General growth law, and from the evolutionary considerations, organisms tend to have such a geometrical form that maximizes the amount of produced biomass for the given conditions. Fraction of nutrients that goes to biomass production is defined by the growth ratio. Among all geometrical forms, a sphere provides the highest value of the growth ratio at the beginning phase of growth and generally the shortest overall cycle time as well (if organisms use the whole possible growth cycle, while some organisms like fission yeast switch to mitosis prematurely). So, first ancient organisms should have a round shape. A spherical shape is also the most restrictive with regard to maximum size[10]. Indeed, spherical microorganisms are among the smallest, and very likely most ancient ones, according to the General growth law. (Of course, other factors shape the geometry of organisms as well.)

So, if we had to decide, which of the two organisms, *S. cerevisiae* or fission yeast *S. pombe*, is less advanced, based on this criterion, we should choose *S. cerevisiae*, as the one that has rounder shape. The obtained inference is opposite to results of studies based on phylogenetic markers. The issue requires further study, of course. Maybe fission yeast originated from an organism *similar* to *S. cerevisiae*, and *S. cerevisiae* appeared later. It is also possible that phylogenetic markers produce biased results. In general, there should be correlation between the advanced growth and division mechanisms and smaller chronological age of species. However, development of more advanced organisms could occur in certain niches, so that more primitive in this regard organisms could develop in parallel or even later. So, less advanced growth and division mechanisms do not necessarily mean a chronologically older organism.

The first ancient organisms should synthesize all components, like RNAs and proteins, at the same rate, which is the simplest and justified from the evolutionary perspective scenario - specialization and diversions generally have to start from something more basic. There are no reasons why the rate of synthesis of different components had to differ at the very first evolutionary stage, since fundamentally synthesis mechanisms are the same for all components. (We can observe such an arrangement in *amoeba*.) So, acquisition of accelerated rate of RNAs synthesis compared to the rate of protein synthesis should be the next evolutionary development. Such a mechanism allowed accelerating biomass synthesis, which gave additional selection advantages. In this regard, both organisms have about a double rate of RNAs synthesis relative to protein synthesis, so that there should be their ancestor, or ancestors, in which all components were synthesized at the same rate.

Double rate of RNA synthesis makes the growth curve computed by the growth equation (5), so to say, more S-shaped, which also means a better expressed inflection point. *Amoeba's* growth curve also has an inflection point, as does the fission yeast, but it is weakly presented. In case of fission yeast, it is much better expressed due to several factors, two of which are cylinder shape and double rate of RNA synthesis compared to rate of protein synthesis. What happened first - acquisition of the double rate or the elongation from a round shape? *S. cerevisiae* apparently has the same division mechanism as *amoeba*[10], which is entirely based on the change of growth ratio. In this situation, elongation would rather decelerate the division because of the more smooth growth curve at the end of growth for such forms, where the division starts. Slightly elongated shape of *S. cerevisiae* is a tribute to other evolutionary pressures. Besides, double rate of RNA synthesis does not require special geometry, while elongated form needs faster rate of biomass production and, optionally, the earlier switching to mitosis, in order to compensate for somewhat slower growth compared to spherical and ellipsoid like forms. So, we may conclude that the accelerated rate of RNAs synthesis appeared evolutionarily before, or at least in parallel with elongation.

What about the cylinder shape? Computations presented in Ref. 10 showed that a cylinder like form growing in one dimension, whose initial length is about two times or more longer than the diameter, has the shortest growth time compared to other elongated forms, like a double cone or a double frustum. Elongation apparently was stimulated by mutation or malfunction that caused premature entering into division. In fact, we still can observe such fission yeast mutants, like *wee*1*Δ* mutant. Such organisms gained a selection advantage, since their reproduction period thus was greatly reduced. Then, such organisms adjusted their division to minimum time, which is the moment when their growth rate reaches maximum, that is the inflection point.

The next (or maybe it was done in parallel) advancement would be to provide the fastest growth by acquisition of the shape that provides the fastest growth among elongated forms, which is the cylinder shape. This is the evolutionary stage where we meet fission yeast. However, *E. coli* and similar organisms could go through similar evolutionary process before fission yeast. It is true that fission yeast *S. pombe* and *E. coli* are very different organisms. However, the optimality of a cylinder form does not depend on size of an organism. So, once we see cylinder like organisms, we may attribute to them, with a high probability, at least two features: (a) they do not go through the whole growth cycle but switch to division at inflection point of the growth curve; (b) they likely have accelerated rate of RNA synthesis compared to the rate of protein synthesis.

With regard to our task, we may conclude with a high probability now that *S. cerevisiae* is an evolutionarily less developed organism than fission yeast. It is also likely that *S. cerevisiae* or similar organism is an ancestor of fission yeast.

Of course, evolutionary applications of the General growth law are by no means restricted to cells and unicellular organisms. In fact, it could be even more revealing to study from this perspective multicellular organisms and plants. For instance, the cone shape of a carrot is very well explained by the General growth law, since a cone shape provides the fastest growth time next to the cylinder shape. The final geometry is always the result of compromise in order to satisfy different criteria. On the same note, we can point to the round shape of fruits, vegetables and berries for which the short growth period is critical, such as native northern plants and berries. Southern plants tend to produce more elongated fruits. The shortest vegetation period for them is not so critical, while having more elongated shape allows acquiring (a) bigger mass, (b) to have long vegetation period and (c) to have less restrictions on the maximum size[10]. So, in this, we also see the influence of the General growth law.

## 5. Conclusion

The application of the General growth law and its mathematical representation the growth equation to the study of microorganisms, illustrated by amoeba and fission yeast examples, allowed to obtain interesting and useful results, both from practical and theoretical perspectives.

First of all, we found that the growth equation, indeed, very adequately describes growth curves of amoeba and fission yeast compared to experimental data. We obtained analytical solution of the growth equation for the fission yeast, which is much more convenient than having a numerical solution.

Second, we convincingly proved that fission yeast, indeed, does not go through the possible growth cycle predefined by the growth equation, but switches to mitosis at inflection point of the growth curve, while amoeba goes almost through the entire possible growth cycle.

Third, based on statistical evaluation, we found that growth curves computed by the growth equation for fission yeast present substantially better approximations of experimental dependencies, compared to known approximations. What makes this result more valuable is that unlike previous studies using mathematical functions just based on the appearance of experimental dependencies, we computed growth curves using real biophysical and geometrical parameters, and only *then* compared the computed growth curves with experiments. All this provides more support for the validity of the General growth law.

Of course, it is almost impossible for biologists to accept an idea that some mathematical equation, but not genetic intricacies, can describe so well growth curves of very different microorganisms, like *amoeba*, fission yeast, its mutant, *S. cerevisiae*, as well as organs, such as livers and liver transplants. Most biologists presently think that biology is infinitely distant from physics, in which, they agree, fundamental laws of nature take place. But not in biology; it is special, it is absolutely different. In fact, biological phenomena are as material as physical ones are. It is all about *matter*, which is governed by fundamental laws of nature. It transforms into different forms, and accordingly different sets of fundamental laws are more prominent for certain matter forms than for the others, but one of the main principles is here to stay: matter is governed by different fundamental laws at different scale levels. Biological phenomena are doomed to be guided by fundamental laws acting at different scale levels. There is no way, in principle, to understand the fundamental causes *why* whale's shape arise from biomolecular mechanisms. Biologists presently know a lot about "how" growth of living organisms happens, but they don't know "why", and, remaining within the biomolecular realm, they unfortunately will never know. Sure, the author of this paper is saying nonconventional things by modern biological standards. However, if we are scientists, then we should not reject such hypotheses upfront, but analyze the submitted proofs and soundness of presented considerations, however controversial they could be in the present state of the discipline. Real breakthroughs always come as heresy to the crowd, and of the higher level they are, the more so they look like.

**Acknowledgements**

The author thanks A. Y. Shestopaloff for editing efforts, discussions and valuable comments, Iva M. Tolic-Norrelykke for helpful discussion, Professor P. H. Pawlowski from Institute of Biochemistry and Biophysics, Warsaw, Professor P. Fantes for continuous support of these studies and comments. Experimental data were courteously provided by S. Baumgartner and I. M. Tolic-Norrelykke from their study done at Max Planck Institute of Molecular Cell Biology and Genetics.

**Appendix A. Finding nutrient influx**

In order to compute the growth curve using equation (5), we need to know specific nutrient influx $k$. It was found in Ref. 10 based on experimental observations and certain considerations supported by experiments, that the rate of nutrient consumption for RNA synthesis in fission yeast is different than for protein synthesis.

The authors of Ref. 37 obtained themselves, and cited other works with similar results, that the rate of synthesis of protein and RNA increases exponentially through the cell cycle. However, the rate of increase can be different. In Ref. 38, the authors obtained the following result for *E. coli*: "The ribosome synthesis rate increases approximately with the square of the growth rate", and also that "in moderate to fast growing bacteria, ... rRNA synthesis per unit protein increases with the square of the cellular growth rate, and ribosomes therefore accumulate in proportion to the growth rate".

According to experimental dependencies obtained in Ref. 11 and other works, discussed in Ref. 10, we can represent the dependence of relative growth of some organism's component's mass on time as an exponential function with base of two, that is

$$m_1(t)/m_{01} = 2^{t\mu} \tag{A.1}$$

where $t$ is time, $m_{01}$ is the beginning mass at point $t = 0$.

If the growth rate for the increase of some other organism's component's mass $m_2$ is double of the previous one, that is $2\mu$, then we can write:

$$m_2(t)/m_{02} = 2^{t(2\mu)} \tag{A.2}$$

or, taking into account equation (A.1), we can write

$$m_2(t)/m_{02} = \left(m_1(t)/m_{01}\right)^2 \tag{A.3}$$

Here, $m_{02}$ is the beginning mass at point $t = 0$. In other words, the relative increase of mass in the second case, when the growth rate is $2\mu$, is the *square* of the relative increase of mass in the first case when the growth rate is $\mu$.

Since for biochemical reactions the law of conservation of mass is fulfilled, the synthesized amount of biomass is equal to the same amount of consumed for that purpose nutrients. Differentiating (A.1) with respect to time, we find nutrient influx $K_p$ required to support protein synthesis:

$$K_p = \frac{C_1}{m_{01}} \times \frac{dm_1(t)}{dt} = C_1 \mu 2^{t\mu} \ln 2 = C_1 \mu \ln 2 \frac{m_1(t)}{m_{01}} \tag{A.4}$$

where $C_1$ is a constant coefficient.

So, the influx required for protein synthesis is proportional to the synthesized mass of protein. If we assume that the relative amount of protein remains constant in the cell, which is usually the case in many cells, equation (A.4) means that the required influx is proportional to cell mass. If the cell density is constant, then we can substitute mass for volume.

As we found above based on results in Refs. 37, 38, for some organisms, the rate of RNA synthesis is double that of protein. (However, this is not always the case. In particular, for *amoeba* rates of protein and RNAs synthesis are the same[10]. So, we can find nutrient influx $K_R$ required for RNAs synthesis by differentiating (A.3), which presents mass increase of substance synthesized at double rate.

$$K_R = \frac{C_2}{m_{02}} \times \frac{dm_2(t)}{dt} = C_2 2\mu 2^{t2\mu} \ln 2 = C_2 2\mu \ln 2 \frac{m_2(t)}{m_{02}} = C_2 2\mu \ln 2 \left(\frac{m_1(t)}{m_{01}}\right)^2 \quad (A.5)$$

where $C_2$ is a constant coefficient.

So, the influx required for RNA synthesis is proportional to the *square* of protein mass. (Note that instead of the base of exponent of two, we could use any other base. The result will be the same.)

The next step is finding the dependence of the total influx on the mass of synthesized components. Note that there is no direct relationship between the consumed nutrients for the synthesis of some components of the cell, and the relative content of these components within the cell. Thus, the relative RNA content may remain the same throughout the cell cycle, although the rate of RNA synthesis may be higher than that of protein. This happens because of different degradation rates for different cell constituents.

It follows from (A.4) that the influx of nutrients that is required for protein synthesis is proportional to the protein mass. Proteins constitute a relatively stable fraction of a cell, and also the largest part of the total cell mass (of up to 5/6 of the dry cell mass[11], so that we can say that nutrient influx required for protein synthesis is also proportional to the cell mass. Accordingly, we may say that the influx of nutrients for synthesis of RNAs, based on (A.5), is proportional to the square of protein mass, and therefore the square of cell mass.

To calculate the total influx, we have to take into account the *relative contents* of protein and RNA. Therefore, we can define the specific influx $k$ required for protein and RNA synthesis as follows.

$$k(m_C) = K(m_C)/S(m_C) = \left(\frac{m_{P0}}{m_{C0}}\left(\frac{m_C}{m_{C0}}\right) + \frac{m_{R0}}{m_{C0}}\left(\frac{m_C}{m_{C0}}\right)^2\right)/S(m_C) \quad (A.6)$$

Here, $m_C$ is the current mass of the cell; $m_{C0}$ is the cell's mass at the beginning; $m_{P0}/m_{C0}$ is the fraction of mass of protein and other cell's components whose rate of synthesis is proportional to the relative cell mass $m_C/m_{C0}$; $m_{R0}/m_{C0}$ is a similar fraction for RNA, whose rate of synthesis is proportional to the square of relative cell mass.

Note that in general, $m_{P0}$ and $m_{R0}$ may vary depending on the phase of the growth cycle, although below we assume them to be constant based on the results of experimental studies, such as the ones presented in Ref. 11.

Let us denote fractions of RNA and protein contents accordingly as $C_r^s$ and $C_p^s$. Assuming that the density is constant, we can define the minimum nutrient influx $K_{\min}$ based on (A.6) as follows[10].

$$K_{\min}(v) = N\left(C_p^s v + C_r^s v^2\right) \quad (A.7)$$

where $N$ is a constant coefficient; $v$ is the fission yeast's relative volume's increase defined as the ratio of the current volume to the initial one, so that $v \geq 1$.

Equation (A.7) does not account for additional nutrients needed to support the transportation and signaling activity, while in order for the synthesis of RNAs and proteins to begin, nutrients first have to be *delivered* to the site of synthesis, which requires energy and nutrients. (For simplicity, we do not consider separately nutrients required for signaling, assuming that they are included into transportation costs.)

In order to find transportation costs, we will use the railway construction analogy, presented in Ref. 10. Let us assume that the weight of materials required to build one unit of railway length is $w$. The price of material, labor and other construction costs per unit of weight is $p_m$. The price of delivering one unit of materials to a one unit distance is $p_t$. In other words, the price of material delivery is proportional to the traveled distance $l$, which is a reasonable assumption. Then, the total price $dP$ of building a new stretch of railway $dl$ at a distance $l$ from the starting point is defined as follows.

$$dP = (w \times dl)p_m + (w \times dl)lp_t = (wp_m + wlp_t)dl \tag{A.8}$$

Solution of this differential equation with respect to $P$, the total price for building a railway stretch of length $l$, is as follows.

$$P = p_m wl + (1/2)p_t wl^2 \tag{A.9}$$

We can see from equation (A.9) that the costs of delivering materials for the railroad construction are proportional to the *square* of the railroad length.

We can similarly consider the "costs" of delivering nutrients to the spot of synthesis in fission yeast in terms of required nutrients, with the only difference that there is no price for nutrients, which means that we should not account for the first term in equations (A.8) and (A.9). The "currency" paid for the nutrients' transportation are nutrients themselves.

Since the diameter of *S. pombe* does not change (or changes little), volume in equation (A.7) is proportional to relative length's increase $L$, which is defined as the ratio of the current length of the cylinder part to the initial one, so that $L \geq 1$.

$$K_{min}(L) = N_1 \left( C_p^s L + C_r^s L^2 \right) \tag{A.10}$$

Here, $N_1$ is a constant. Now, we have to add the transportation costs to influx required for the protein and RNAs synthesis. Similarly to the second term in equation (A.9), we can write the following equation for the total influx $K$ that takes into account transportation costs.

$$dK(L) = C_t K_{min}(L)dL \tag{A.11}$$

Here, $C_t$ is a constant coefficient. Equation (A.11) is a mathematical expression that we need to add transportation costs to nutrients required for synthesis of biomass and maintenance. Substituting equation (A.10) into equation (A.11) and solving it, we obtain the following.

$$K(L) = N_1 C_t \left( (1/2)C_p^s L^2 + (1/3)C_r^s L^3 \right) = A \left( C_p L^2 + C_r L^3 \right) \tag{A.12}$$

where $A = N_1 C_t / 2$, while

$$C_p = C_p^s \; ; \; C_r = (2/3)C_r^s \tag{A.13}$$

Note that the ratio of total nutrient influxes required for the RNA and protein synthesis, as it is defined by (A.13), has been changed, so that in order to find the contents of these substances, we have to make a translation from the nutrient influx to the actual RNA content using (A.13).

In order to match time units between the computed growth curve and experimental data, we will have to use a scaling constant, so that all the above constants will include this scaling coefficient as a common multiplier, which, of course, won't change the shape of the computed growth curve.

It is not always possible to analytically solve equation (A.11), especially when a growing organism has complex geometry. In this case, one has to use a more general form of the growth equation that incorporates transportation (and signaling) costs[10].

The nutrient influx required for two- and three-dimensional growth can be found similarly[10]. Assuming that the rate of RNAs synthesis is double of that for proteins, for a disk-like organism with a constant height growing in two dimensions, the total influx is defined as

$$K(v) = \left( C_p v^{3/2} + C_r v^{5/2} \right) \tag{A.14}$$

For a three-dimensional form proportionally increasing in three dimensions, we have:

$$K(v) = \left(C_p(v)^{4/3} + C_r(v)^{7/3}\right) \tag{A.15}$$

## Appendix B. Analytical solution of the growth equation for the fission yeast

Here, we find solution of equation (9).

$$p\pi r^2 l_b dL = A(C_p L^2 + C_r L^3) \frac{(2/3)R(E-L)}{((4/3)R+L)(2R+E)} dt \tag{B.1}$$

We take into account that $l = l_b L$ and consequently $dl = l_b dL$; $R = r/l_b$. Surface area $S$ in equation (5) will be cancelled by the same term in the denominator of specific nutrient influx, since $k = K/S$.

Separating variables and doing appropriate transformations, we obtain the following differential equation.

$$\left(\frac{3p\pi R l_b^3 (2R+E)}{2C_p}\right) \frac{(4/3)R + L}{L^2(1+(C_r/C_p)L)(E-L)} dL = A dt \tag{B.2}$$

Note that both the left and right parts of equation (B.2) have mass dimensions.

Let us denote $C = C_r/C_p$ and $B = 3p\pi\pi R_b^3(2R+E)/(2C_p)$, which are constants. For simplicity, we first find the integral

$$\int_{L_b}^{L} \frac{(4/3)R + L}{L^2(1+CL)(E-L)} dL \tag{B.3}$$

and then multiply the found solution by $B$.

We find the integral of the left part as a sum of integrals. In order to do that, we use the presentation with unknown values of $d, f, g, h$ to be found.

$$\frac{(4/3)R + L}{L^2(1+CL)(E-L)} = \frac{Ld+f}{L^2} + \frac{g}{1+CL} + \frac{h}{E-L} \tag{B.4}$$

Transforming the right part of equation (B.4) to a common denominator, we obtain the following.

$$\frac{L^3(hC - g - dC) + L^2(dCE - d - fC + gE + h) + L(dE - f + fCE) + fE}{L^2(1+CL)(E-L)} \tag{B.5}$$

Equating coefficients for the appropriate powers of $L$ in equation (B.5) and the right part of equation (B.4), we obtain the following system of four equations with four unknown values.

$$hC - g - dC = 0$$
$$dCE - d - fC + gE + h = 0$$
$$dE - f + fCE = 1$$
$$fE = (4/3)R \tag{B.6}$$

Solving this system, we find.

$$f = \frac{4R}{3E}; \quad d = \frac{1+f-fCE}{E}; \quad h = \frac{d+fC}{CE+1}; \quad g = C(h-d) \tag{B.7}$$

Substituting the right part of equation (B.4) into equation (B.2), and doing integration, we finally obtain

$$t = \frac{B}{A}\left(f\left(\frac{1}{L_b} - \frac{1}{L}\right) + d\ln\left(\frac{L}{L_b}\right) + \frac{g}{C}\ln\left(\frac{1+CL}{1+CL_b}\right) + h\ln\left(\frac{E-L_b}{E-L}\right)\right) \tag{B.8}$$

Note that the last term in equation (B.8) provides a vertical asymptote, that is when $L \to E$, $t \to \infty$. Accordingly, the reverse function $L(t)$, which, from a geometrical perspective, is a symmetrical reflection of function $t(L)$ defined by equation (B.8) relative to the line $L = t$, has a term with horizontal asymptote.